\documentclass[sigconf, nonacm= true]{acmart}

\usepackage[center]{subfigure}
\usepackage{booktabs}
\newcommand{\para}[1]{\smallskip\noindent\textbf{#1}}

\AtBeginDocument{%
  \providecommand\BibTeX{{%
    \normalfont B\kern-0.5em{\scshape i\kern-0.25em b}\kern-0.8em\TeX}}}

\copyrightyear{2022} 
\acmYear{2022} 
\setcopyright{rightsretained} 
\acmConference[WebSci '22]{14th ACM Web Science Conference 2022}{June 26--29, 2022}{Barcelona, Spain}
\acmBooktitle{14th ACM Web Science Conference 2022 (WebSci '22), June 26--29, 2022, Barcelona, Spain}\acmDOI{10.1145/3501247.3531583}
\acmISBN{978-1-4503-9191-7/22/06}

\begin{document}

%%%%%%%%%%%%%%%%%%%%%%%%%%%%%%%%%%%%%%%%%%%%%%%%%%%%%%%%%
% TITLE
%%%%%%%%%%%%%%%%%%%%%%%%%%%%%%%%%%%%%%%%%%%%%%%%%%%%%%%%%
\title{Link recommendations: Their impact on network structure and minorities}

% \title{How recommendations shape the structure of networks and the impact on minorities}

% \title{Feedback loops in network-based recommender systems}

%%%%%%%%%%%%%%%%%%%%%%%%%%%%%%%%%%%%%%%%%%%%%%%%%%%%%%%%%
% AUTHORS
%%%%%%%%%%%%%%%%%%%%%%%%%%%%%%%%%%%%%%%%%%%%%%%%%%%%%%%%%
\author{Antonio Ferrara}

\affiliation{
  \institution{GESIS - Leibniz Institute for the Social Sciences}
  \city{Cologne}
  \country{Germany}}
\affiliation{%
  \institution{RWTH Aachen University}
  \city{Aachen}
  \country{Germany}
}  
\email{antonio.ferrara@gesis.org}

\author{Lisette Espín-Noboa}

\affiliation{%
  \institution{Complexity Science Hub}
  \city{Vienna}
  \country{Austria}}
\affiliation{%
  \institution{Central European University}
  \city{Vienna}
  \country{Austria}}

\email{espin@csh.ac.at}

\author{Fariba Karimi}

\affiliation{%
  \institution{Complexity Science Hub}
  \city{Vienna}
  \country{Austria}}

\email{karimi@csh.ac.at}

\author{Claudia Wagner}

\affiliation{%
  \institution{GESIS - Leibniz Institute for the Social Sciences}
  \city{Cologne}
  \country{Germany}}
\affiliation{%
  \institution{RWTH Aachen University}
  \city{Aachen}
  \country{Germany}
  }
  \affiliation{%
  \institution{Complexity Science Hub}
  \city{Vienna}
  \country{Austria}}
\email{claudia.wagner@gesis.org}

%%\affiliation{%
%%  \institution{Institute for Clarity in Documentation}
%%  \streetaddress{P.O. Box 1212}
%%  \city{Dublin}
%%  \state{Ohio}
%%  \country{USA}
%%  \postcode{43017-6221}
%%}

\renewcommand{\shortauthors}{Ferrara et al.}

%%%%%%%%%%%%%%%%%%%%%%%%%%%%%%%%%%%%%%%%%%%%%%%%%%%%%%%%%
% ABSTRACT
%%%%%%%%%%%%%%%%%%%%%%%%%%%%%%%%%%%%%%%%%%%%%%%%%%%%%%%%%
\begin{abstract}
Network-based people recommendation algorithms are widely employed on the Web to suggest new connections in social media or professional platforms.
While such recommendations bring people together, the feedback loop between the algorithms and the changes in network structure may exacerbate social biases. 
These biases include rich-get-richer effects, filter bubbles, and polarization.
However, social networks are diverse complex systems and recommendations may affect them differently, depending on their structural properties.
In this work, we explore five people recommendation algorithms by systematically applying them over time to different synthetic networks.
In particular, we measure to what extent these recommendations change the structure of bi-populated networks and show how these changes affect the minority group. %s in bi-populated networks.
\let\thefootnote\relax\footnotetext{Please cite the WebSci'22 version of this paper. 

Antonio Ferrara, Lisette Espín-Noboa, Fariba Karimi, and Claudia Wagner. 2022. Link recommendations: Their impact on network structure and minorities. In 14th ACM Web Science Conference 2022 (WebSci ’22), June 26–29, 2022, Barcelona, Spain. ACM, New York, NY, USA, 11 pages. https://doi.org/10.1145/3501247.3531583}

Our systematic experimentation helps to better understand when link recommendation algorithms are beneficial or harmful to minority groups in social networks. In particular, our findings suggest that, while all algorithms tend to close triangles and increase cohesion, all algorithms except Node2Vec are prone to favor and suggest nodes with high in-degree. Furthermore, we found that, especially when both classes are heterophilic, recommendation algorithms can reduce the visibility of minorities.

\end{abstract}

\begin{CCSXML}
<ccs2012>
   <concept>
       <concept_id>10002951.10003260.10003282.10003292</concept_id>
       <concept_desc>Information systems~Social networks</concept_desc>
       <concept_significance>500</concept_significance>
       </concept>
   <concept>
       <concept_id>10002951.10003317.10003347.10003350</concept_id>
       <concept_desc>Information systems~Recommender systems</concept_desc>
       <concept_significance>500</concept_significance>
       </concept>
 </ccs2012>
\end{CCSXML}

\ccsdesc[500]{Information systems~Social networks}
\ccsdesc[500]{Information systems~Recommender systems}
%\ccsdesc[500]{Computer systems organization~Embedded systems}

\keywords{Recommendation algorithms, friendship recommendations, network science, social networks, homophily, preferential attachment.}
\maketitle

%%%%%%%%%%%%%%%%%%%%%%%%%%%%%%%%%%%%%%%%%%%%%%%%%%%%%%%%%
% INTRO
%%%%%%%%%%%%%%%%%%%%%%%%%%%%%%%%%%%%%%%%%%%%%%%%%%%%%%%%%

\section{Introduction}

Social networks are the infrastructure of our social and professional life. 
They impact, among others, our cooperation~\cite{Fowler5334}, our health~\cite{Christakis2013}, and our social perceptions~\cite{lee2019homophily}. 
The structure of modern online social networks is however not only shaped by well-studied social mechanisms (such as homophily or preferential attachment), but it is also affected by people recommender systems, complex algorithms that suggest new connections among social network users. 
How do these algorithms affect the structure of social networks over time? What are the consequences for different groups? 
In this paper, we aim to shed light on these questions. 
%We focus on social networks in which nodes can be described by a binary attribute (e.g. male or not male).

\para{Problem:}
% The introduction of the Web 2.0 created a participatory culture among online users.
%Since the introduction of the Web 2.0, users may create, share or like content while being part of many social networks. 
%This participatory culture has allowed users to discover and interact with many more people than in previous years.
%Since it is impossible to keep track of everything that is being generated online, recommender systems facilitate this seeking of information by matching users' preferences and their current connections.
Previous work has shown that recommendation algorithms are prone to reinforcing popularity bias~\cite{abdollahpouri2019popularity}. 
A further subtle problem is that by matching users' preferences, these algorithms often lead to the formation of filter bubbles \cite{chitra2020analyzing}, echo chambers \cite{baumann2020modeling}, and polarization \cite{dandekar2013biased}.
In recent years, much attention has been paid to understanding when, and to what extent, such biases are being amplified. 
As an example,~\cite{fabbri2020effect} and~\cite{espin2022ineq} have studied the correlation between network structure and the output of ranking algorithms in social networks. While these studies highlight that \textit{homophily}---the tendency to connect to similar others---and  \textit{preferential attachment}---the tendency to connect to those that
are already well-connected---are important structural factors that impact the visibility of nodes in algorithmic rankings, they do not compare effects over time. Feedback loops, instead, have been studied in \cite{su2016effect} and \cite{stoica2018algorithmic}, where they respectively analyze ``rich-get-richer'' and ``glass ceiling'' effects. Recently, also \cite{fabbri2021exposure} and \cite{cinus2021effect} have focused on feedback loops and long term effects of people recommender systems. The former analyzes inequalities in the exposure of minorities and the latter focuses on polarization and echo chambers. Our study integrates this body of research by providing a systematic analysis of how homophily and minority size relate to structural properties of the network and visibility of groups.

%They have found that \textit{homophily}---the tendency to connect to similar others---is the main driver of disparities such as over- or under-representation of minorities in top-k ranks.
%Additionally, the \textit{activity} of nodes---or tendency of creating out-links---has been found important for amplifying the visibility or representation of minorities in the rank, especially when one group is more active than the other~\cite{espin2022ineq}. \textit{Preferential attachment}, on the other hand, known as the tendency to connect to popular people, shows little influence on these disparities between groups, but it amplifies inequality at a node level (i.e., only a few important nodes are being recommended)~\cite{espin2022ineq}.
%While these studies represent great advances towards more fair, accountable, and transparent recommendation algorithms, they do not study the feedback loop effects on the network structure over time. 

\para{Approach:}
We systematically compare five recommendation algorithms and apply their recommendations to several synthetic networks. 
We focus on scale-free directed networks with adjustable homophily and minority group size~\cite{espin2022ineq} and quantify the global changes in network structure, as well as the changes in connectivity for the minority group over time.
In particular, we assess whether certain types of links are created more often than others and whether the network becomes more cohesive or segregated. 
Similarly, we verify when, and at what rates, these algorithms put minorities at disadvantage by measuring the changes in their \textit{visibility}, here defined as the fraction of minorities among the top most important nodes, based on their algorithmic ranking.
To this end, we formulate the following research questions that will guide our analysis throughout this paper.

% -algorithmic fairness in networks and difference with `tabular data`  fairness (the features are the links and the network structure)

% -Studies on homophily and group visibility /(fairness), studies on node degree /individual fairness

% -connecting recommendation algorithms on network with algorithmic fairness

% -in which case and which algorithms preserve the network structure and preserve the types of connections (homophily) between the classes + this would be desiderable in a WYSIWYG fairness worldview since under this worldview we assume that the initial network before recommendation represent the natural and correct way of connecting nodes

% -Instead a WAE worldview assumes that the belonging to one or the other classes should not influence the relevance and the characteristics of a node. We will hence study when and which recommendation algorithms can favor a balance between the node importance of the two classes

% \lesp{Suggestion for RQs: 
% RQ1: How do recommendation algorithms affect the structure of the network and the visibility of minorities? (Fig 3 and fig Gini-CC).
% RQ2: To what extent is the change in visibility due to homophily? (fig 4 and 5) 
% RQ3: Is the change in visibility proportional to the size of the minority? (new fig not yet in paper)}

% \para{Research questions:}
\begin{itemize}
    \item RQ1: How do recommendation algorithms affect the structure of the network and the visibility of minorities?
    \item RQ2: To what extent is the change in visibility due to homophily?
    \item RQ3: Is the change in visibility inversely proportional to the size of the minority or proportional to the in-group links within the minority?
    
    % \item RQ1: How do recommendation algorithms affect the visibility of the
% minority group in the network? In particular, under which conditions of homophily and group size are recommendation algorithms increasing or decreasing the visibility of minorities? 

% To what extent does the interplay between homophily,preferential attachment, and the size of the minority affectthe visibility of minority nodes in the network1
    % \item RQ2: Which changes in the network structure are responsible for the variations of the visibility of the minorities? In particular, to what extent the over- and under-representation of minorities at the top of algorithmic rankings is related to the changes in inter- and intra-class connections? 

    %To what extent does the interplay between homophily, preferential attachment, and the size of the minority affect the visibility of minority nodes in the network?

    % \item RQ1: To what extent do recommendation algorithms change the structure of networks? Are recommendation algorithms equally changing the structure of groups in the network? (i.e., the changes made to the network equally affect majority and minority groups). 
    % \item RQ2: Under which conditions are recommendation algorithms increasing/decreasing the visibility or importance of minority nodes in the network?
    %How do recommendation algorithms are changing the structure of networks? and how the connections between the different classes are affected?
    % \item RQ2: Are recommendation algorithms increasing disparities among different classes? And in particular disadvantaged minorities are even more penalized?
\end{itemize}

% \para{Approach:} xxx
% - description of our experiment

\para{Contributions:} 
Our contributions are the following: (1) We demonstrate that networks become more cohesive over time throughout multiple recommendations. However, the rate at which this cohesiveness gets stronger depends on the algorithm. (2) Not all algorithms suffer from the popularity bias problem, which means that certain algorithms may diversify their recommendations. (3) The visibility of the minority group gets affected differently depending on three main components: the algorithm, the initial conditions of homophily in the network, and the size of the minority group.

%Our systematic study serves to understand and to explain the long-term effects of recommendations in social networks. 
Moreover, our study sheds light on the weaknesses of algorithms under the initial conditions of network structure and can be used as key factors to improve recommendations, where necessary.
% extends previous work by adding global explanations for each algorithm given certain types of data.

% - description of findings
% - general impact
% - considerations

\begin{figure}
    \centering
    \includegraphics[width=0.4\textwidth]{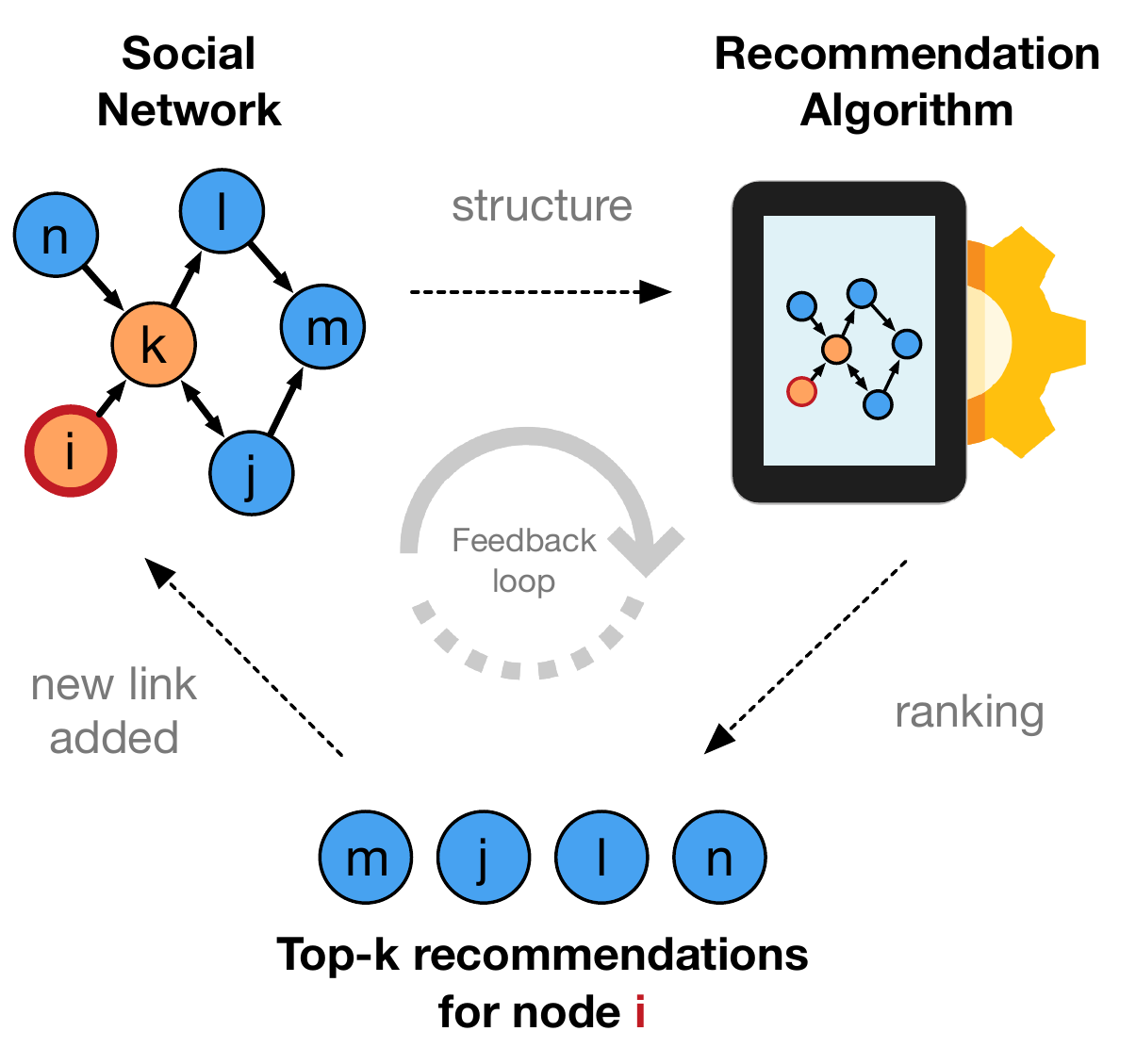}
    \caption{The recommendation cycle: 
    A network-based recommendation algorithm uses the local or global structure of the network to recommend for each node $i$ the top-k best matches with whom node $i$ may want to connect. If node $i$ accepts the recommendations,  the structure of the network changes. This creates a feedback loop since the new structure is pre-processed by the algorithm to infer new recommendations.
    }
    \label{fig:example}
\end{figure}

%%%%%%%%%%%%%%%%%%%%%%%%%%%%%%%%%%%%%%%%%%%%%%%%%%%%%%%
% RELATED WORK
%%%%%%%%%%%%%%%%%%%%%%%%%%%%%%%%%%%%%%%%%%%%%%%%%%%%%%%%%

\section{Related Work}

%In networks, nodes are not all equal and might differ in terms of structural properties and connections. These structural properties can themselves constitute source of biases and inequalities. For example, by affecting the visibility of the nodes or in terms of ability to spread or receive information. 

The related work is organized in two parts. First, we introduce the relevant literature on the mechanisms that drive the existence of biases in network structure. % structural biases in networks. 
Then, we focus on the creation of new ties from link recommendations. In particular, we highlight the effects of recommendation algorithms on the network structure and the visibility of minorities.%, intended as their position in an algorithmic ranking. 

\para{Biases in network structure and related consequences:}  
The rich-get-richer or Matthew effect \cite{merton1988matthew} is one of the first mechanisms of edge formation discovered by sociologists to explain cumulative advantages in real-world networks.
%that have been discovered %and that drive the formation of edges 
%with the consequent creation of biases in the structure by reinforcing cumulative advantages. % \cite{merton1988matthew} introduced the Matthew effect and the cumulative advantage behavior, a tendency to give more credit to already famous people. 
From the network perspective, the Matthew effect operates through the preferential attachment mechanism, that is the tendency of nodes to attach preferentially to those that are already well-connected~\cite{barabasi1999emergence}. %\lesp{Here you can also cite blackwell1973ferguson, sornette1998multiplicative. They also model/reinforce the rich get richer effect. Something along these lines: "The rich-get-richer effect, also known as preferential attachment [], cumulative advantage [], P\'olya Urn [], and multiplicative process [], ..."}
%Homophily, the tendency to connect to similar others \cite{mcpherson2001birds}, propensity of similar nodes to connect with each other and how, in certain cases, this phenomenon can create strong divides in groups is studied in ~\cite{mcpherson2001birds}. 
This mechanism of edge formation and other structural characteristics may impact the visibility and importance of nodes, and thus, create and enlarge inequalities. For example,~\cite{avin2015homophily} and~\cite{karimi2018homophily} propose mathematical models that integrate preferential attachment and homophily (the tendency to connect to similar others~\cite{mcpherson2001birds}) to explain the emergence of the ``glass ceiling'' effect in social networks. Glass ceiling, as defined by the US Federal Commission, is ``the unseen, yet unbreakable barrier that keeps minorities and women from rising to the upper rungs of the corporate ladder, regardless of their qualifications or achievements''. Studies on glass ceiling are expanded in~\cite{nilizadeh2016twitter}, where the authors consider the effect of the perceived gender on the visibility of users on Twitter. In particular, they reveal how users perceived as women are hampered from attaining equal visibility. Furthermore,~\cite{karimi2018homophily} shad light on how homophily can put minority groups at disadvantage by restricting their ability to establish links with the majority group and by limiting their access to information. Recently,~\cite{tsioutsiouliklis2021fairness} observed that PageRank~\cite{page1999pagerank} might unfairly allocate importance scores to different classes, and proposed alternative fair versions of the algorithm.

Our work is built upon this body of literature and integrates social biases in feedback loops of people recommendations. In particular, we analyze the tendencies of groups to connect to each other, how these tendencies or mechanisms of edge formation affect the recommendations, and ultimately how these recommendations affect the structure of networks and the visibility of minorities.

\begin{figure*}[t!]
    \centering
    \includegraphics[width=\textwidth]{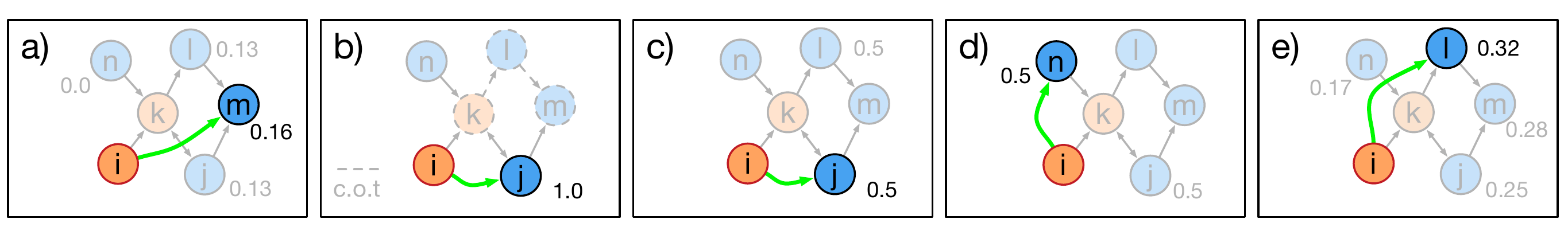}
    \caption{Recommendation algorithms: Given the network in Figure \ref{fig:example}, here we explain the recommendation suggested to node $i$ by each of our algorithms of interest. 
    %Note that all algorithms are class or color agnostic and their recommendations are based on different topological strategies.
    % a) Personalized PageRank performs random walks with restart. Personalization for node $i$ implies that the random walk restarts  always from node $i$. The algorithm recommends the most visited nodes by the random walks: $i\rightarrow m$. 
    a) Personalized PageRank recommends the most visited nodes by performing random walks that restart from node $i$: $i\rightarrow m$.
    b) Who-to-follow builds the so called ``circle of trust'' (c.o.t.: nodes k, l, and m) for $i$ and recommends nodes that are followed by the nodes in the c.o.t.: $i\rightarrow j$.
    c) Two-hops recommends nodes at a distance 2: $i\rightarrow j$.
    d) Common-followed suggests nodes with a similar set of out-links: $i\rightarrow n$.
    e) Node2Vec projects nodes into an euclidean space and recommends those with similar embeddings: $i\rightarrow l$.
    %b) Node2Vec projects nodes in an euclidean space, creating a vectorial representation for each node. Then, the algorithm recommends nodes that are close in terms of cosine similarity in the euclidean space. Relative distances between nodes in the network tend to be preserved in the euclidean space, hence pathwise close nodes in the network tends to be recommended: $i\rightarrow l$.
     The values next to each node are the scores returned by each algorithm. The larger the value, the more important the node for $i$.
    }
    \label{fig:algorithms}
\end{figure*}

%%here in the pdf the space is different but I can't understand why
\para{Effects of recommender systems on networks:}
\cite{su2016effect} analyzes the ``rich-get-richer'' phenomenon through social recommendations. In particular, they study how the ``Who-to-Follow'' algorithm affects the structure of the follower network on Twitter. % the changes on the follower network on Twitter's network structure due to the introduction of its ``Who to Follow'' recommendations. 
They found that most popular users profited substantially more than average from the user suggestions. They attributed this ``rich-get-richer'' effect to various factors, including the mismatch between users (being recommended proportional to their degree), and the baseline growth rate of users (whose asymptotic behavior is instead sub-linear in the degree).
Users' centrality and clustering coefficient may also vary depending on the recommendation algorithm in ``Social-Blue'', an internal social networking site at IBM~\cite{daly2010network}.
% shows that users' centrality and clustering may vary greatly depending on the recommendation algorithm in ``Social-Blue'', an internal social networking site at IBM.
%analyzes the effect of different recommendation algorithms on users in ``Social-Blue'', the IBM social network. They showed that, depending on the recommendation algorithm employed, the effects on the nodes of the network may vary greatly with respect to centrality and clustering measures. 
Similar effects have been found in Tumblr and Flickr, two social media platforms, where recommendations favor popular and well-connected nodes, and at the same time limit the growth of the diameter of the network~\cite{aiello2017evolution}.

In addition to these topological effects, social recommendations may also exacerbate the under-representation of certain demographic groups in the network.
% For instance, \cite{stoica2018algorithmic,espin2022ineq}, show xxx
% in networks with homophily and preferential attachment.
% studies the presence of ``glass ceiling'' effects in link recommendations on Instagram. %the Instagram social network. 
% and relative social recommendations. 
% In particular, they distinguish glass ceiling effects due to organic growth and recommendation models, showing that, while algorithms do not create disparity out of nowhere, they can worsen pre-existing inequalities in networks with homophily and preferential attachment.
For instance, \cite{stoica2018algorithmic,fabbri2020effect,espin2022ineq} 
%studies people recommender systems in social networks and 
show how the visibility of minorities can be amplified or mitigated by different levels of homophily within groups when using recommendation algorithms on scale-free networks.
% people recommender systems in networks with preferential attachment.
%each subgroup. 
% Close to our work, \cite{fabbri2021exposure} studies the long-term effects of multiple rounds of recommendations.
% Recently, long-term effects of multiple rounds of recommendations have been studied to understand their implications~\cite{fabbri2021exposure}. 
These inequalities have been also studied over time but only recently. \cite{fabbri2021exposure} suggests that while the homophily level of the minority affects the speed of the growth of their disparate exposure, the relative size of the minority affects the magnitude of this effect.

%by including multiple rounds of recommendations and considering long-term effects which we aim in this work too.

% shows that a minority group, if homophilic enough, can get
% a disproportionate advantage in exposure from all link recommenders. Instead, when it is heterophilic, it gets underexposed. Moreover, while the homophily level of the minority affects the speed of the growth of the disparate exposure,
% the relative size of the minority affects the magnitude of the
% effect. Finally, link recommenders strengthen exposure inequalities at the individual level, exacerbating the“rich-getricher” effect: this happens for both the minority and the majority class and independently of their level of homophily.

One of the main differences between this body of research and our work is that we vary homophily systematically. This allows us to better understand the relationship between the initial homophily of the network and the long-term effects of the recommendations. In particular, to what extent they change network structure and the visibility of minorities over time.
%Furthermore, our analysis connects the changes in the visibility of the minorities with the changes in network structure. 
Moreover, we study Node2Vec~\cite{DBLP:journals/corr/GroverL16}, a more recent algorithm used to generate link recommendations through node embeddings.
% a more recent method based on node embeddings, in particular 

%Recommendations algorithms and opinion-dynamics simulation models have been combined in \cite{cinus2021effect} to study echo chambers and polarization effects in social networks. Their experimentation shows that people recommendation systems can lead to a significant increase in echo chambers and that this happens independent from the initial homophily in the network. Different from our work, they consider undirected networks and symmetric homophily levels, without focusing on the impact of recommendation algorithms on minority and majority classes. 

%%\lesp{how is our work related (or not) to this body of literature?}

%% AF not sure if we should include this part and in this way. 
%% \para{Fairness in recommender systems:} xxx \lesp{here all about definition of fairness and interventions), filter bubbles (related to preferences/similarities), and their respective interventions wrt fairness -not necessarily on on networks-. Focusing on networks is the gap we want to bridge. So, we can conclude this paragraph with how we would do it.}. \cite{chen2020bias}.  \cite{zehlike2021fairness}. \cite{pitoura2021fairness}. \lesp{how is our work related (or not) to this body of literature?}

%%%%%%%%%%%%%%%%%%%%%%%%%%%%%%%%%%%%%%%%%%%%%%%%%%%%%%%%%
% METHODS
%%%%%%%%%%%%%%%%%%%%%%%%%%%%%%%%%%%%%%%%%%%%%%%%%%%%%%%%%

\section{Methods}

\subsection{Directed networks}
\label{sec:net}

%We employ the DPAH model which allows to generate scale-free bi-populated directed networks with adjustable homophily (for each group), minority size, node activity, and edge density~\cite{espin2022ineq}. % we put this later in the experimental setup
We consider attributed directed networks of the following form: let $G=(V,E,C)$ be a node-attributed graph where $V=\{v_1,...,v_n\}$ is a set of $n$ nodes, $E \subseteq  V \times V$ is a set of $e$ unweighted directed edges, and $C:V \longrightarrow \{0,1\}$ is a function that maps each node $v_i$ into its group (or class) membership $c_i$. For the sake of simplicity we focus on binary group membership (e.g., black/white or male/non-male). The function $C$, hence, divides the nodes into two groups, a minority, called $m$, and a majority, called $M$. We refer to the fraction of the minority group in the network as $f_m$.  

Further definitions, peculiar to the synthetic network generation model employed, are provided in Section \ref{sec:setup}.

\subsection{Recommendation algorithms}
\label{sec:recalg}
In this section, we define the five recommendation algorithms of interest. 
All algorithms are class agnostic which means that their recommendations are solely based on topology.
Note as well that for each node $v_i \in V$, the recommendation algorithm suggests a ranked list of $k$ nodes that $v_i$ is not yet connected with. The ranked list is sorted in descending order in terms of relevance scores according to each algorithm. In the case of ties, where multiple nodes are equally relevant, nodes are chosen randomly.
We refer the reader to Section \ref{sec:setup} for the details on the configuration of hyper-parameters for each algorithm.

%\begin{figure*}[b]
%\includegraphics[width=0.30\textwidth]{Cfexplain.png} 
%\caption{\textbf{Example.}  Explain \lesp{here all methods}}
%\label{fig:example}
%\end{figure*}

\para{Personalized PageRank (PPR)}: It is an extension of PageRank to rank nodes in a network from the perspective of a seed node~\cite{page1999pagerank}.
In principle, random walks are performed and restarted at the origin (or seed node) multiple times to update the importance score of all nodes, see Figure~\ref{fig:algorithms}(a). % to rank all other nodes from the perspective of the origin. %seed node's point of view. 
% We compute the PPR vector $\pi_i$ for each node $v_i \in V$.
% do this for all nodes separately, and use their rankings to suggest them the node with highest (personalized) PageRank (see \ref{fig:example}). % to each node the highest ( % top to follow the higher (not already connected) node ranked from the seed perspective.
We compute the PPR vector $\pi_i$ with respect to each node $v_i \in V$ as follows:
\begin{equation}
    \pi_i^T = (1-\alpha) e_i^T+ \alpha \pi_i^TW
\end{equation}
where $\alpha$ is the probability of following links, $e_i$ denotes the personalized one-hot vector\footnote{$(e_i)_i=1$ and $(e_i)_j=0$, $\forall j \neq i$}, $W$ is the transition matrix inferred from $G$ and $T$ represents the transpose operator. % and $|\pi_i|=n$. 
The ranking score given to node $v_j$ is then the $j^{th}$ component of $\pi_i$.

\para{Who-to-follow (WTF)}: 
% This algorithm was proposed by Twitter to suggest new people to follow~\cite{gupta2013wtf}.
% It is based on SALSA~\cite{lempel2001salsa} and Personalized PageRank~\cite{jeh2003scaling}.
This algorithm, proposed by Twitter~\cite{gupta2013wtf}, suggests users who are followed by people that are similar to the one getting the recommendation, see Figure~\ref{fig:algorithms}(b).
% a user $v_i$ is likely to follow those who are followed by users that are similar to $v_i$.
For each user $v_i$, the algorithm looks for its \textit{circle of trust}, which is the result of an egocentric random walk (similar to personalized PageRank~\cite{jeh2003scaling}). 
Then, based on this circle-of-trust $COT_i$, WTF ranks (using the SALSA algorithm~\cite{lempel2001salsa}) users that are not yet friends with $v_i$ but are connected through the circle of trust $\pi^{out}_{COT_i}$. 
\begin{equation}
    WTF_i = SALSA(COT_i, \pi^{out}_{COT_i})
\end{equation}
%Then, we take the top-10 of these (recommended) users, and add up the counter of being selected as a recommendation to each of them. This is done for every node $v_i$ in the network. At the end, the rank of each node encodes the \textit{number of times a user was suggested as a recommendation} across all nodes in the network. Thus, the WTF score for each node is defined as follows:
% where $SALSA$ refers to the top-k users the SALSA algorithm recommends to node $i$ based on its circle of trust 

\para{Two-hops (2H)}: 
This algorithm follows the intuition behind \textit{friends-of-friends}. 
In directed networks, the 2H algorithm recommends nodes $v_j$ that are at a distance 2 from node $v_i$, see Figure~\ref{fig:algorithms}(c).
% Two hops is a variation of common-followed. Here, the set of out-neighbors for $v_j$, $\Gamma^{out}_j$, the set of in-neighbors for $v_j$, $\Gamma^{in}_j$, that is defined as the set of nodes that have a link directed to $v_j$. 
The more such paths, the more likely the recommendation.
Calling $\Gamma^{out}_a$ the set of nodes that $v_a$ points towards (i.e., out-links), and $\Gamma^{in}_a$ the set of nodes pointing to $v_a$ (i.e., in-links), we define the 2H score function as the number of possible paths of length 2 from $v_i$ to $v_j$:
\begin{equation}\label{eq:2H}
\mathrm{2H}(v_i,v_j) \coloneqq \vert \Gamma^{out}_i \cap \Gamma^{in}_j \lvert 
\end{equation}

\begin{figure*}[t!]
    \centering
    \includegraphics[width=0.99\textwidth]{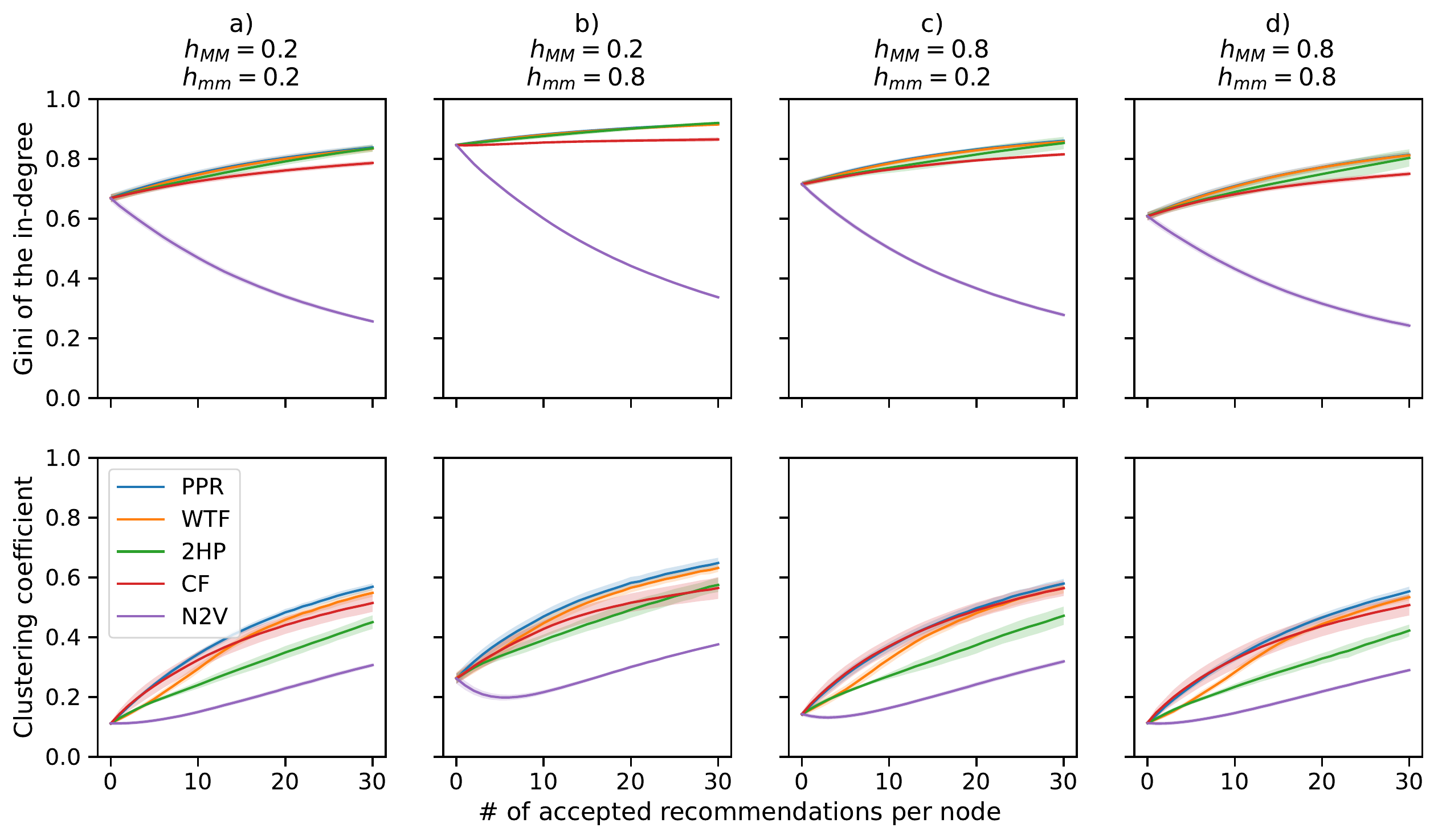}
    \caption{The evolution of network structure % the Gini coefficient of the in-degree distribution and the clustering coefficient 
    for different recommendation algorithms and different values of homophily in the initial network. One can see that, regardless of the type of network (columns), all algorithms except N2V have similar effects on the Gini coefficient of the in-degree distribution (top row) and the clustering coefficient (bottom row). Surprisingly, N2V reduces the Gini coefficient of the in-degree distribution over time (x-axis) and increases the global clustering coefficient at lower rates compared to the other algorithms.}
    \label{fig:netstructure}
\end{figure*}

\para{Common-followed (CF)}: We extend the common neighbors approach~\cite{liben2007link}, which is based on the idea that two nodes $v_i$ and $v_j$ are more likely to connect to each other if they have multiple friends in common. In the context of directed networks, the \textit{common-followed} algorithm will recommend node $v_j$ to node $v_i$ if they follow partially or fully the same set of nodes, see Figure~\ref{fig:algorithms}(d). Then, the algorithm ranks all nodes $v_j$ based on the number of common-followed nodes with $v_i$.
Let $\Gamma^{out}_a$ be the set of nodes that $v_a$ follows. We define the set of common-followed nodes between $v_i$ and $v_j$ as:
% based on the idea that two nodes $v_i$ and $v_j$ are more likely to connect in the future if their sets of neighbors $\Gamma_i$ and $\Gamma_j$ have a large overlap. In the context of social networks, this approach follows the natural intuition that such nodes $v_i$ and $v_j$ represent users and if these two users have many friends in common, then, they are more likely to come into contact and become friends themselves. The common \textit{neighbors} score function has been used in the context of link prediction for undirected networks~\cite{liben2007link}. Given the score function, the recommendation for a fixed node $v_i$ is then computed by finding another node $v_j$ with largest score\footnote{In case of ties---when more than one node possess the same score---we choose any of these nodes at random.}. % that is not already a neighbor of $x$ (breaking randomly in case of ties). 
% In an analogous way has been done in \cite{stoica2018algorithmic} for the Adamic-Adar coefficient. We have further extended the common neighbors for directed networks as follows. 
%Let $\Gamma^{out}_i$ be the set of nodes that $v_i$ follows (i.e., there is a link from $v_i$ to them, also called from now on out-neighbors). We will refer to this approach as common followed. 
\begin{equation}\label{eq:CF}
\mathrm{CF}(v_i,v_j) \coloneqq \vert \Gamma^{out}_i \cap \Gamma^{out}_j \lvert 
\end{equation}
% where $\Gamma^{out}_a$ is the set of nodes that $v_a$ follows (i.e., there is a link from $v_a$ to each $v_b \in \Gamma^{out}_a$).

\para{Node2Vec (N2V)}: A popular embedding algorithm that maps nodes to a low-dimensional space of features, by maximizing the likelihood of preserving nodes' neighborhoods~\cite{DBLP:journals/corr/GroverL16}. It has been used for link prediction by evaluating the cosine similarity between nodes in the embedding space, see Figure~\ref{fig:algorithms}(e). Here we use N2V to recommend to each node $v_i$ the most similar node in the embedding space, according to cosine similarity of the embedded vectors. Calling respectively $v^p_i$ and $v^p_j$ the embedded vector projections for $v_i$ and $v_j$, the cosine similarity between these projections is defined as: 
\begin{equation}\label{eq:cosine sim}
\mathrm{CosineSim}(v^p_i,v^p_j) \coloneqq \frac{v^p_i \cdot v^p_j}{\rVert  v^p_i \lVert \rVert v^p_j \lVert }
\end{equation}

%%%%%%%%%%%%%%%%%%%%%%%%%%%%%%%%%%%%%%%%%%%%%%%%%%%%%%%%%
% EXPERIMENTS
%%%%%%%%%%%%%%%%%%%%%%%%%%%%%%%%%%%%%%%%%%%%%%%%%%%%%%%%%

\subsection{Experiments setup}
\label{sec:setup}
Here we describe the networks employed in our experiments and explain how the recommendation algorithms are iteratively used to recommend new connections among nodes. 
%We assume that only the top-1 recommendation is accepted. % with an equal chance (in our case 100\%). 
%This decision is motivated by the fact that the employed acceptance policy plays only a marginal role in shaping the network~\cite{fabbri2021exposure, cinus2021effect}.

\para{Synthetic networks:} In order to systematically create networks as defined in Section \ref{sec:net}, we employed the DPAH model~\cite{espin2022ineq}. This model allows to generate scale-free bi-populated directed networks with adjustable homophily (for each group), minority size, node activity, and edge density.
%describe here how it works and the parameters
DPAH is a growth model that generates networks as follows. First, $n$ nodes are created and randomly assigned to one of two groups based on the 
%labeled according to the
fraction of minorities $f_m$. Then, the following steps are repeated until the desired edge density $d$ is fulfilled.
A source node $v_i$ is drawn from a power-law distribution, modeled through the activity parameters $\gamma_M$ and $\gamma_m$ for the majority and the minority group, respectively. A target node $v_j$ is drawn with a probability that is proportional to the product of its in-degree and the pair-wise homophily between the source and the target node. Lastly, a directed edge from $v_i$ to $v_j$ is created.
% In formula, calling $v_i$ the source node and $v_j$ the target node, 
Thus, the probability of creating a link from $v_i$ to $v_j$ is defined as:
\begin{equation}
    \mathbb{P}(v_i \rightarrow v_j) = \frac{h_{ij}k_{j}^{in}}{\sum _{l=1}^{n}h_{il}k_{l}^{in}}
\end{equation}
where $k_{j}^{in}$ is the in-degree of $v_j$, and $h_{ij}$ is the homophily between $v_i$ and $v_j$ and it is determined by their class membership.

In this work, we systematically modify the homophily within groups and the size of the minority, leaving the variation of node activity and edge density for a further study. In particular, in order to measure the influence of algorithms (RQ1) and homophily (RQ2) in the recommendations, we generate $4$ networks for each combination of homophily parameters $h_{mm},h_{MM}\in\{0.0, 0.1, \dots, 1.0\}$ ($h_{mM}$ and $h_{Mm}$ are defined as $1 - h_{mm}$ and $1 - h_{MM}$, respectively) and fix the number of nodes $n=1000$, the size of the minority $f_m=0.3$, the node activity $\gamma_M=\gamma_m=2.5$ and the edge density $d=0.03$. We further adjust the size of the minority $f_m\in\{0.1,0.2,0.3,0.4\}$ to measure its influence in the visibility of minorities (RQ3).

\para{Recommendation:} Given an initial network $G$, we apply a recommendation algorithm $R$ to suggest to each node $v_i$ a node $v_j$ to connect with. Then, we create a direct link $v_i \rightarrow v_j$ for each top-1 of these recommendations. By doing so, in what we call ``one step'', we create a new out-link for each node $v_i$. This decision is motivated by the fact that the employed acceptance policy plays only a marginal role in shaping the network~\cite{fabbri2021exposure, cinus2021effect}.
Then, for every addition, we remove a random out-link. This is a procedure previously employed in the literature, for example in~\cite{cinus2021effect}. One of the main reasons for this choice is to prevent a significant increase in the edge density of the network. The evaluation metrics considered in Section~\ref{sec:metrics} are sensible to edge density. By removing a link every time a new one is created we ensure to keep the density constant on every step and make sure that the changes %in the metrics studied 
are due to the recommendations and not %due 
to an increase in the total amount of connections. The link removal procedure is also grounded on the social theory for which people exhibit a finite communication capacity and, thus, they have a limit on the number of ties that they can maintain active in time~\cite{dunbar1992neocortex,miritello2013limited}. 

We repeat the above procedure $30$ times to simulate an equal amount of recommendations per node.

\para{Hyper-parameters:} For PPR, we set the probability of following links to $\alpha=0.85$, as suggested by Brin and Page \cite{brin1998anatomy} and widely used in many applications. In N2V, we use the default values for the dimensions of the embedding space ${dimensions}=64$, the number of visited nodes in each random walk ${walk\_length}=10$, and the number of random walks to be generated from each node in the graph ${num\_walks}=200$. For WTF, we constrain the circle of trust to include only the top-10 nodes.

\para{Additional assumption:} We assume that the recommendations of different algorithms are similarly relevant, as our goal is not to evaluate which algorithm performs better in terms of utility metrics, but rather to study their effects on the structure and their impact on the visibility of the minorities (see Section \ref{sec:metrics}).

\subsection{Evaluation metrics}
\label{sec:metrics}

We use the global \textit{clustering coefficient}~\cite{fagiolo2007clustering} of the network and the \textit{Gini coefficient}~\cite{gini1912variabilita} of the in-degree distribution as proxies of network structure, and the fraction of minorities among the most important nodes as \textit{visibility}. We measure these metrics before and after each round of recommendations to verify whether certain types of networks change these metrics faster or slower and by how much.

\para{Clustering coefficient:} This metric allows to verify whether the recommendations are making the network more cohesive by closing more triangles. The clustering coefficient of node $v_i$ is defined as:
\begin{equation}
c_{v_i}=\frac{2}{\operatorname{deg}^{t o t}(v_i)\left(\operatorname{deg}^{t o t}(v_i)-1\right)-2 \operatorname{deg}^ \leftrightarrow(v_i)} T(v_i)
\end{equation}
where $T(v_i)$ is the number of directed triangles through node $v_i$, $\operatorname{deg}^{t o t}(v_i)$ is the sum of in-degree and out-degree of $v_i$, and $\operatorname{deg}^{\leftrightarrow}(v_i)$ is the reciprocal degree of $v_i$. The global clustering coefficient of the network is then obtained by taking the mean across all nodes: $c=1/n\sum_{i=1}^n c_{v_i}$.

\para{Gini coefficient of the in-degree distribution:} Popularity bias is a well-known issue reinforced by certain recommendation algorithms~\cite{abdollahpouri2019popularity}. The Gini coefficient~\cite{gini1912variabilita} allows us to demonstrate whether this bias is exacerbated by the algorithms regardless of the initial conditions of the network structure, or whether certain types of networks are exempt from this bias. The Gini coefficient of the in-degree distribution $\pi^{in}$, sorted in ascending order, is defined as follows:
\begin{equation}
Gini(\pi^{in})=\frac{\sum_{i=1}^{n} (2i- n -1)\pi^{in}_i}{n\sum_{i=1}^{n} \pi^{in}_i}
\end{equation}
% where $\pi^{in}$ represents the in-degree values for all nodes, indexed in a ascending order.

The higher the Gini coefficient, the more skewed or unequal the in-degree distribution across all nodes. %non-decreasing order.

\para{Visibility of the minority group:} First, we measure the importance of nodes by computing their PageRank~\cite{page1999pagerank}. Then, out of the top-10\% highest-scored nodes, we measure the fraction of nodes that belong to the minority group and refer to this fraction as the visibility of the minority group $\hat{f_m}$. 
We use the relative visibility $\hat{f_m^*}=\hat{f_m}-f_m$ to verify how far the visibility of the minority is from statistical parity~\cite{dwork2012fairness} before the recommendations.
Finally, we measure the change in visibility by computing $\hat{f_m}$ after and before the recommendations to verify whether the minority group is gaining or losing visibility:
% w.r.t., their size in the network (i.e., analogous to statistical parity~\cite{dwork2012fairness}).
\begin{equation}
    \delta_{f_m} = \hat{f_m}(after) - \hat{f_m}(before)
\end{equation}

\para{In-group links:} We also look at the fraction of links within groups to see what type of edges are being recommended more often by the algorithms.
The in-group link ratio for group $a$ is defined as:
\begin{equation}
    I_a = \frac{e_{aa}}{e_{aa}+e_{ab}}
\end{equation} where $a,b \in \{m,M\}$ and $a\neq b$.
% [define here the metrics used to validate and inspect the results eg clustering coefficient, fraction of links etc]

%%%%%%%%%%%%%%%%%%%%%%%%%%%%%%%%%%%%%%%%%%%%%%%%%%%%%%%%%
% SYNTHETIC
%%%%%%%%%%%%%%%%%%%%%%%%%%%%%%%%%%%%%%%%%%%%%%%%%%%%%%%%%

\section{Results}

Here we address our three research questions and present the results obtained after applying the recommendation algorithms iteratively to the simulated directed networks described in Sections \ref{sec:recalg} and \ref{sec:setup}, respectively. First, we show the consequences of these recommendations on the structure of the network and on the visibility of the minority group (RQ1). Second, we explain the changes in structure and visibility as a function of homophily (RQ2). Third, we further investigate the role of the size of the minority group and in-group links in the effects of the recommendations (RQ3).

%the effects that recommendations have on the network structure and how these structural changes relate to the nodes visibility.

\begin{figure}[t!]
    \centering
    \includegraphics[width=0.48\textwidth]{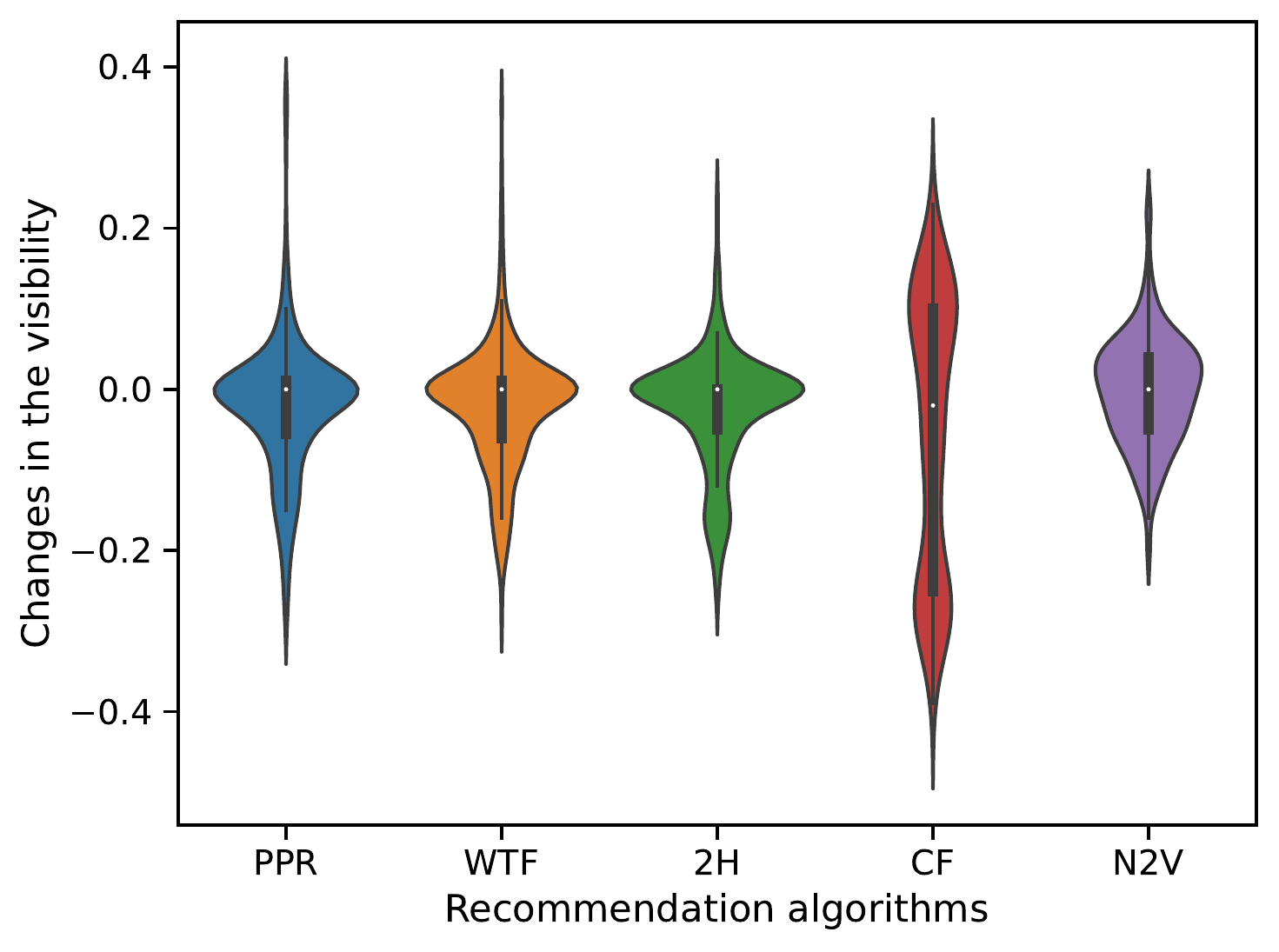}
    \caption{Changes in the visibility of minorities for different recommendation algorithms. After 30 recommendations for each node (in networks with different values of homophily and fixed minority size $f_m=0.3$), we see that all algorithms may increase (positive change), decrease (negative change) or keep constant (zero change) the visibility of the minority group. In particular, PPR, WTF and 2H mostly keep the visibility of minorities unchanged. However, their tails are asymmetric and denser in the negative direction, in correspondence with a decrease in visibility. CF maintains the initial visibility of minorities in a few cases, and otherwise it may drastically change the visibility in both directions. 
    N2V generates less extreme changes in both directions. % with less variations.% as with the other algorithms. 
    %here also fm 0.3
    }
    \label{fig:vio_diff}
\end{figure}

% \subsection{How recommendation algorithms are affecting the visibility of minority nodes in the network? In particular, under which conditions of homophily and group sizes are recommendation algorithms increasing or decreasing the rankings of minorities? }

\subsection{RQ1: How do recommendation algorithms affect the structure of the network and the visibility of minorities?}
\label{Resultsrq1}

\begin{figure*}[t!]
    \subfigure[Before recommendation]{\includegraphics[height=0.25\textwidth, width=0.34\textwidth]{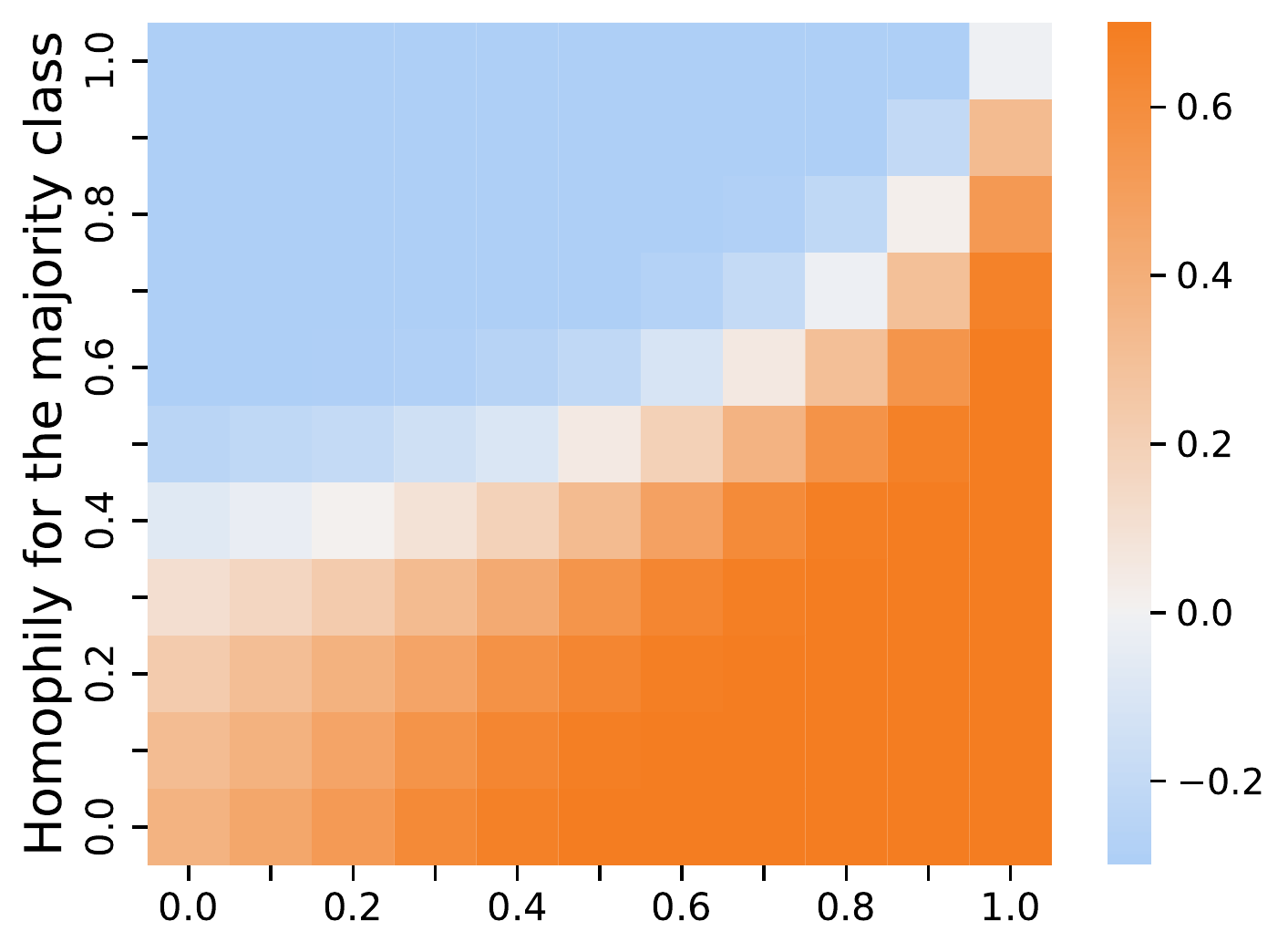} \label{fig:hv_before}}
    \subfigure[Variation after PPR]{\includegraphics[height=0.25\textwidth, width=0.31\textwidth]{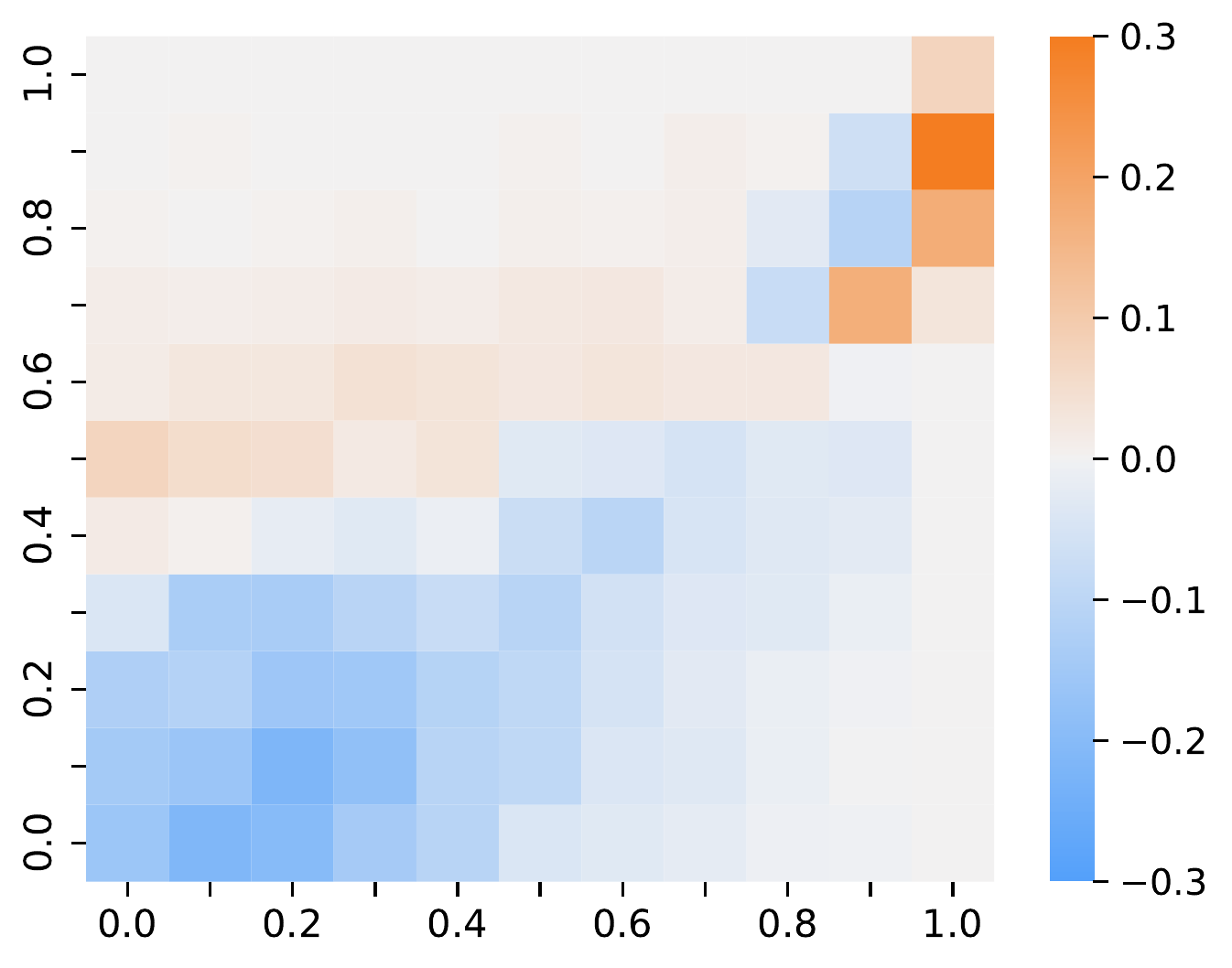} \label{fig:hv_ppr}}
    \subfigure[Variation after WTF]{\includegraphics[height=0.25\textwidth, width=0.31\textwidth]{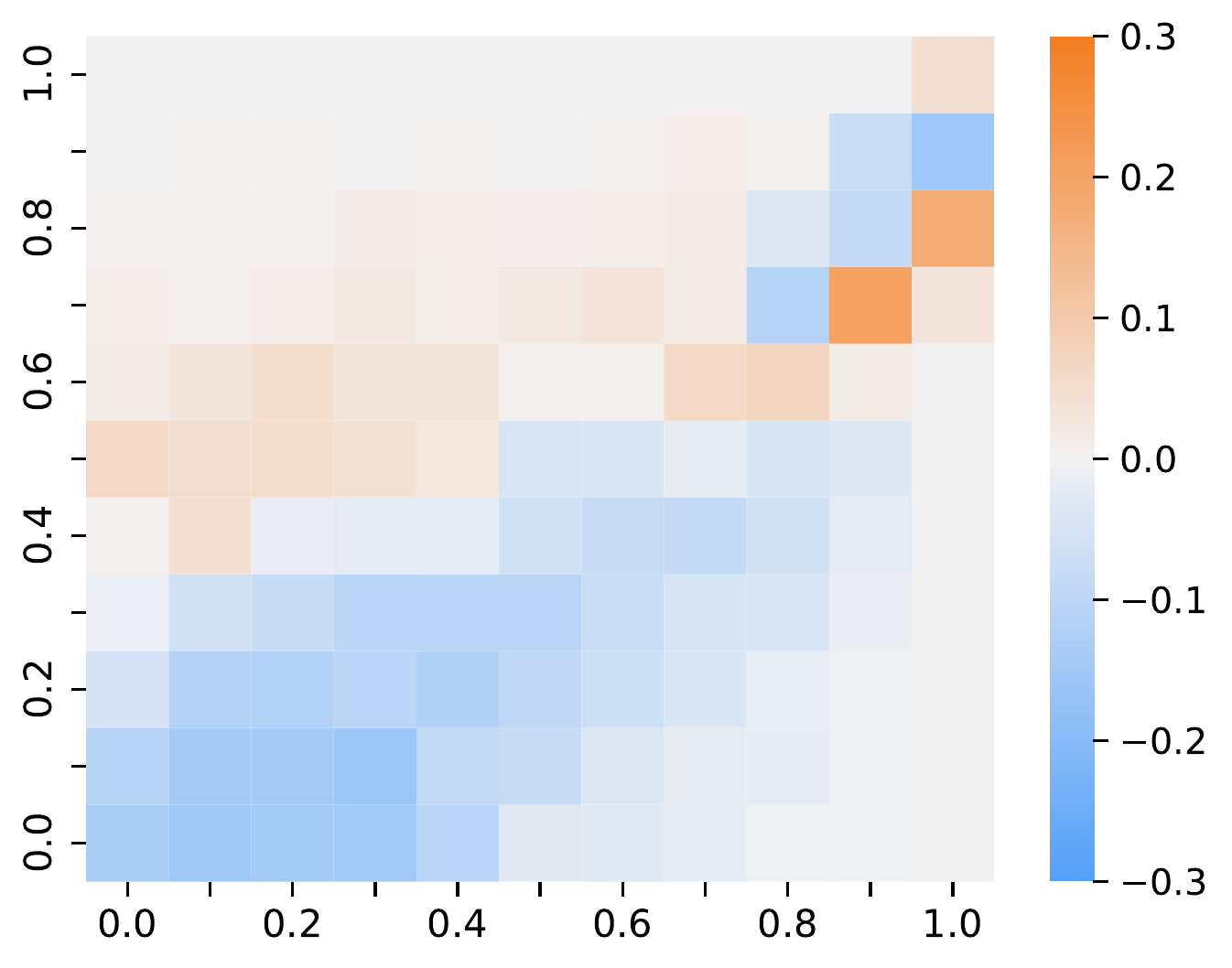} \label{fig:hv_wtf}}
    \subfigure[Variation after 2H]{\includegraphics[height=0.26\textwidth, width=0.34\textwidth]{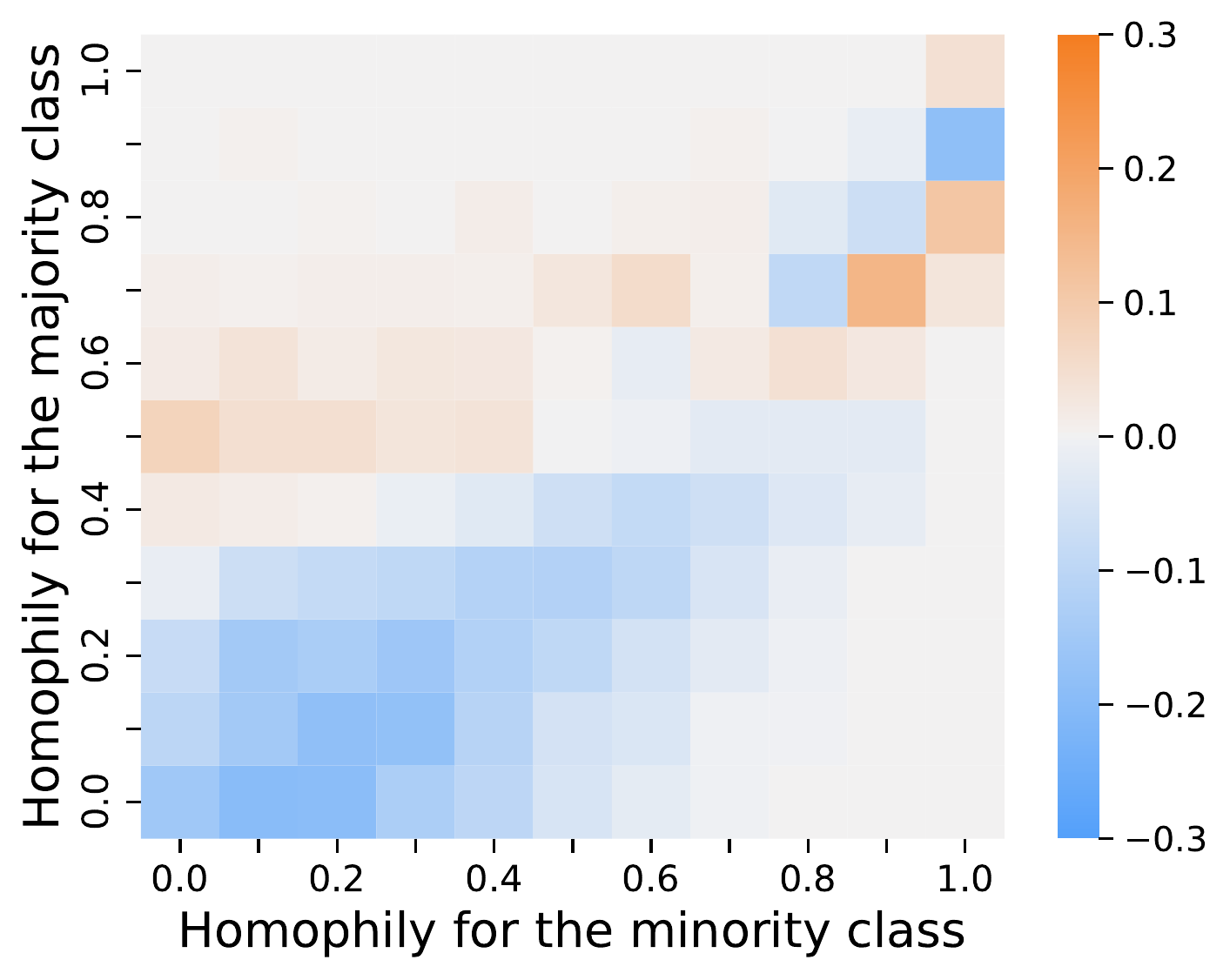} \label{fig:hv_2H}}
    \subfigure[Variation after CF]{\includegraphics[height=0.26\textwidth, width=0.31\textwidth]{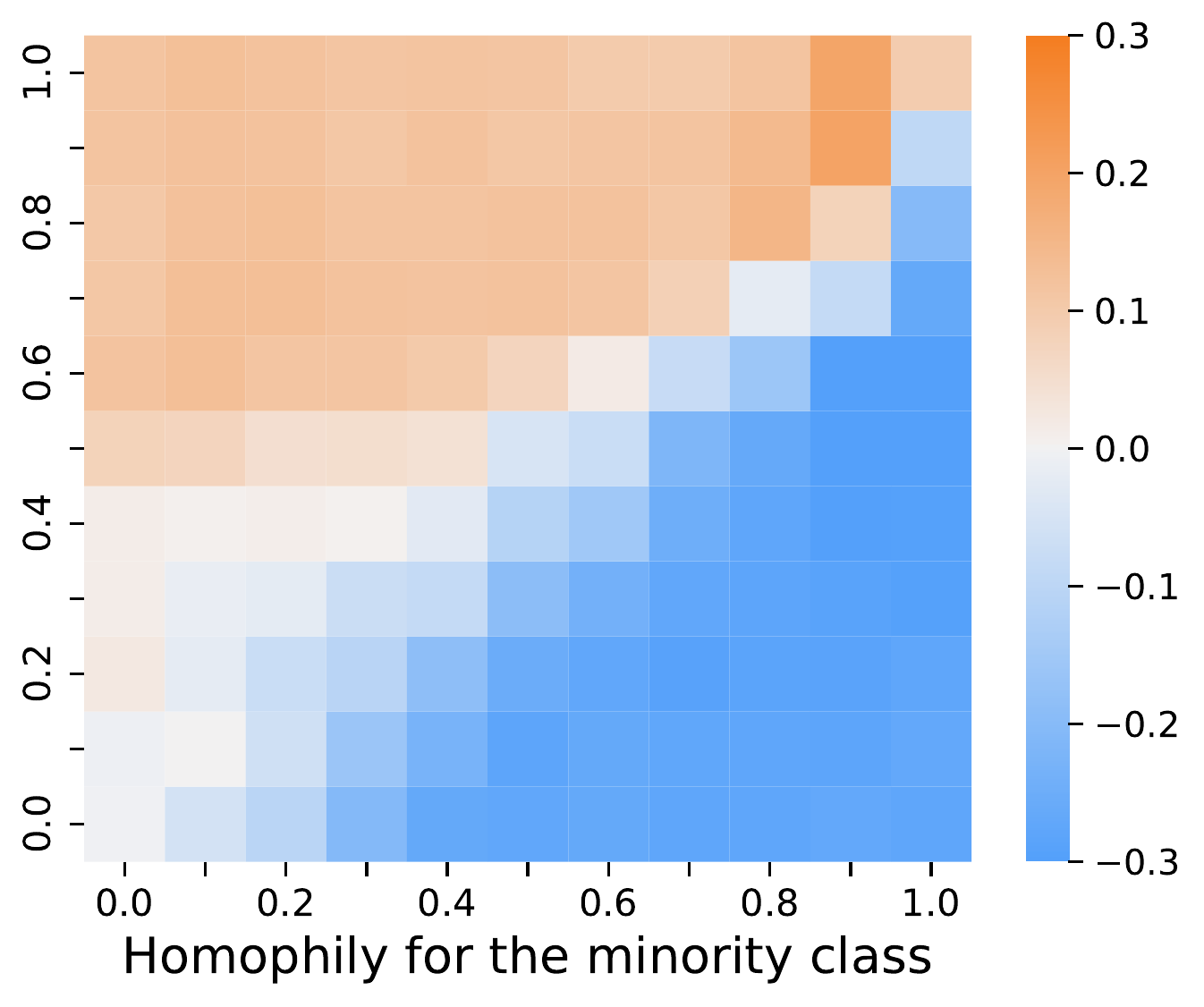} \label{fig:hv_CF}}
    \subfigure[Variation after N2V]{\includegraphics[height=0.26\textwidth, width=0.31\textwidth]{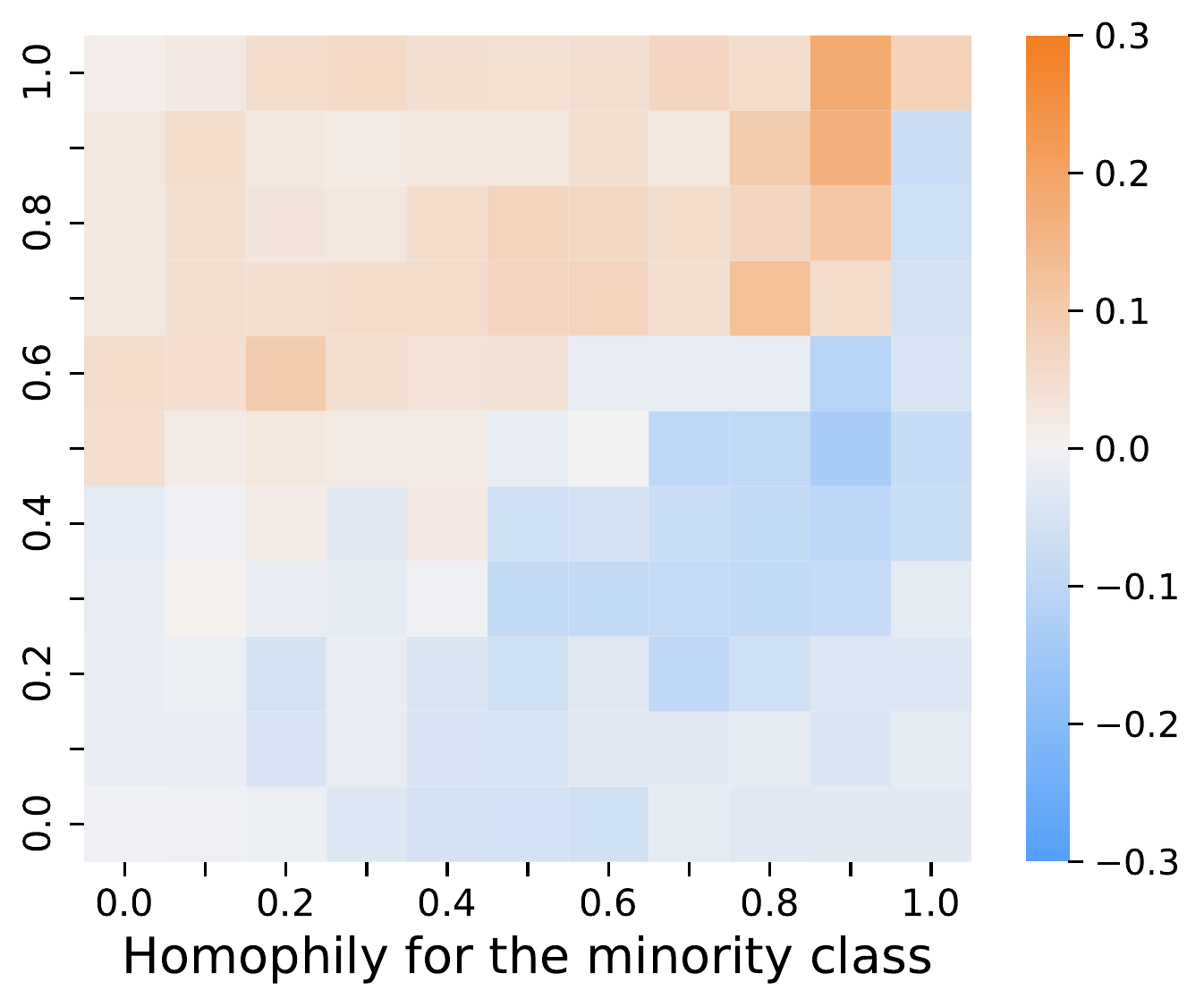} \label{fig:hv_N2V}}
    \caption{Visibility of the minority group as a function of homophily. 
    %Fraction of minority in top 10\% ranked by PageRank. 
    % Squares in 
    % before and after 30 recommendations
    Heatmaps show the visibility of the minority group before and after the recommendations for different algorithms and different combinations of homophily within the majority (y-axis) and the minority (x-axis) groups.
    %types of networks with initial values of homophily within the majority (y-axis) and within the minority group (x-axis). %parameters on the $x$ and $y$ axis. 
    % In (a), 
    The visibility of the minority group is measured by the fraction of minorities in the top-10\% of nodes ranked by their PageRank. %w.r.t. the fraction of minorities in the network. 
    %of the minority in the top 10\% of nodes ranked by their PageRank. . 
    In (a), colors show the relative visibility of the minority group w.r.t., the fraction of minorities in the network $f_m=0.3$ before the recommendations. Positive visibility means that the minority is over-represented (orange), and negative visibility means that the minority is under-represented or the majority is over-represented (blue). Zero visibility refers to those cases where the top rank does not include any node from the minority group.
    In (b-f), colors represent the variation in the visibility due to different recommendation algorithms. For PPR, WTF and 2H one can see that the minority loses more visibility than the majority (especially in the heterophilic regime), while CF and N2V show more symmetric effects on the visibility of the minority and majority. %counteract over-representation of minorities in heteropilic regimes, while CF and N2V strongly and slightly contrast both over- and under-representations of minorities, respectively
    % Here fm 0.3
    Notice that the homophily values shown in the x- and y-axis of all plots represent the initial levels of homophily in the network before the recommendations.
    }
    \label{fig:hv}
\end{figure*}

%%% HERE STRUCTURE
\para{Changes in network structure:} To address this question, we first assess the changes in network structure in terms of global clustering coefficient and Gini coefficient of the in-degree distribution. The idea is to verify whether the algorithms (while connecting people together) make the network more cohesive and whether popularity bias increases at the same rate for all algorithms.
Figure \ref{fig:netstructure} shows the results for both metrics (top/bottom) on different types of networks (columns) across multiple rounds of recommendations (x-axis). Note that the x-axis reflects the iteration or step of recommendation, e.g., at step=20, each algorithm has independently recommended 20 connections to each node in the network.
First, we see that overall, the evolution of these metrics is consistent across types of networks (columns) and recommendation algorithms (colors). 
Second, all recommendation algorithms increase the clustering coefficient of the network which means that the networks are becoming more cohesive as more triangles are getting closed. However, the rate at which this clustering increases, differs across algorithms, especially for N2V which, surprisingly, is the slowest.
Third, we corroborate that PPR, WTF, 2H and CF reinforce the popularity bias issue since the Gini increases over time.
This means that these algorithms make popular people (in terms of high in-degree) more popular.
The exception is N2V, which after several recommendations makes the in-degree distribution less skewed (i.e., the recommendations are more diverse). One possible explanation is that similarity in the embedding space is only partially sensible to popularity bias.

%% HERE VISBILITY
\para{Changes in the visibility:} Now, we explore to what extent each recommendation algorithm changes the visibility of the minority group after several recommendations.
%To address this question, we explore how the structure of the network changes after 30 recommendations for each node in the network.
We show the results in Figure~\ref{fig:vio_diff}. 
%we see that the how the visibility of the minority changes for different recommendation algorithms. 
Each violin refers to one algorithm and the distribution of the violin represents the variation across a multiplicity of networks with different initial homophily values, fixed number of nodes and minority size (see Section~\ref{sec:setup} for details). 
PPR, WTF and 2H show similar patterns: they have median close to zero but denser tails in the negative direction. 
This indicates that these algorithms mostly keep the visibility of the minority unchanged, but, in certain cases, they decrease this visibility. % significantly decrease the visibility of the minorities in certain types of networks. 
CF shows the opposite behavior. First, it keeps the visibility unchanged for a few cases, but most of the time it drastically changes this visibility in either direction. %presents a bi-modal distribution, increasing or decreasing severely the visibility of the minority at the top of the ranks. 
% However, the intensity is larger against the minority due to higher negative values. 
Among all, N2V reveals more symmetric and smaller effects. 
Summarizing, Figure~\ref{fig:vio_diff} suggests that four out of five algorithms are more prone to keep the visibility of the minority unchanged. Nevertheless, in certain regimes (explored next in RQ2) this visibility can be increased or reduced depending on the levels of homophily. 
% suggests that for 4 out of 5 algorithms, iterative rounds of recommendations reduced the visibility of minorities more. 

\subsection{RQ2: To what extent is the change in visibility due to homophily?} 
\label{sec:rq2}
% Now, the following questions might be raised: under which condition of initial homophily the visibility of minority is reduced? And in the situations in which they are penalized by the recommendations algorithms, they were already in a disadvantage or instead they were over-represented at the top of the rankings? 

To understand how the initial levels of homophily in the network affect the recommendations, we compare the visibility of the minority before and after the recommendations for each algorithm, see Figure~\ref{fig:hv}. We control for the number of nodes and the fraction of minorities by keeping them fixed, and vary homophily values (see Section~\ref{sec:setup} for more details).
As defined in Section~\ref{sec:metrics}, visibility measures the fraction of nodes that belong to a particular group and make it to the top-10\% of the rank. This rank reflects the importance of nodes in the network and it is assessed through their PageRank~\cite{page1999pagerank}. 

\para{Visibility before the recommendations:} Figure~\ref{fig:hv_before} shows the relative visibility of the minority before the recommendations. White regions (neutral visibility) represent statistical parity~\cite{dwork2012fairness}, in which the fraction of the minority in the top-10\% is equal to the fraction of minority populating the whole network. Orange regions (positive visibility) represent higher amount of minority nodes at the top of the rank compared to the statistical parity condition. Blue regions (negative visibility), instead, represent under-representation of minorities in top ranks.
We see that the minority is over-represented mostly when the majority is heterophilic $h_{MM}<0.5$ or when the minorities are more homophilic than the majorities $h_{mm}>h_{MM}$.

\para{Changes in the visibility after the recommendations:} Figures \ref{fig:hv_ppr} to ~\ref{fig:hv_N2V} show the change in visibility after $30$ recommendations per node. 
%of the minorities is changing due to the new connection introduced by the recommendations. 
A positive change (orange) indicates that the visibility of the minority increased after the recommendations (relative to the initial visibility they had before the recommendations). Actual values in each cell denote the magnitude of this change.
Conversely, a negative change (blue) indicates that the majority increased its visibility at the cost of reducing the visibility of the minority. No changes (white) indicate that the visibility did not vary across time.
At first glance, we see that the visibility gets affected differently depending on the algorithm and the initial values of homophily. We further notice that there are slightly more blue than orange regions in almost all plots (i.e., the majority increases its visibility more often than the minority across all regimes). 

Among all the algorithms CF produces the strongest changes, while N2V is the most balanced. PPR, WTF and 2H, on the other hand, show a similar behavior. They penalize minorities especially in the heterophilic regimes for both classes, i.e., $h_{**}<0.5$, bottom-left corners of Figures ~\ref{fig:hv_ppr} to ~\ref{fig:hv_2H}. 

Furthermore, when only one group is homophilic, PPR, WTF and 2H do not change the initial over-representation of the homophilic group, see top-left and bottom-right corners in Figures~\ref{fig:hv_ppr} to ~\ref{fig:hv_2H}.

\begin{figure*}[t!]
    \centering
    \includegraphics[width=0.99\textwidth]{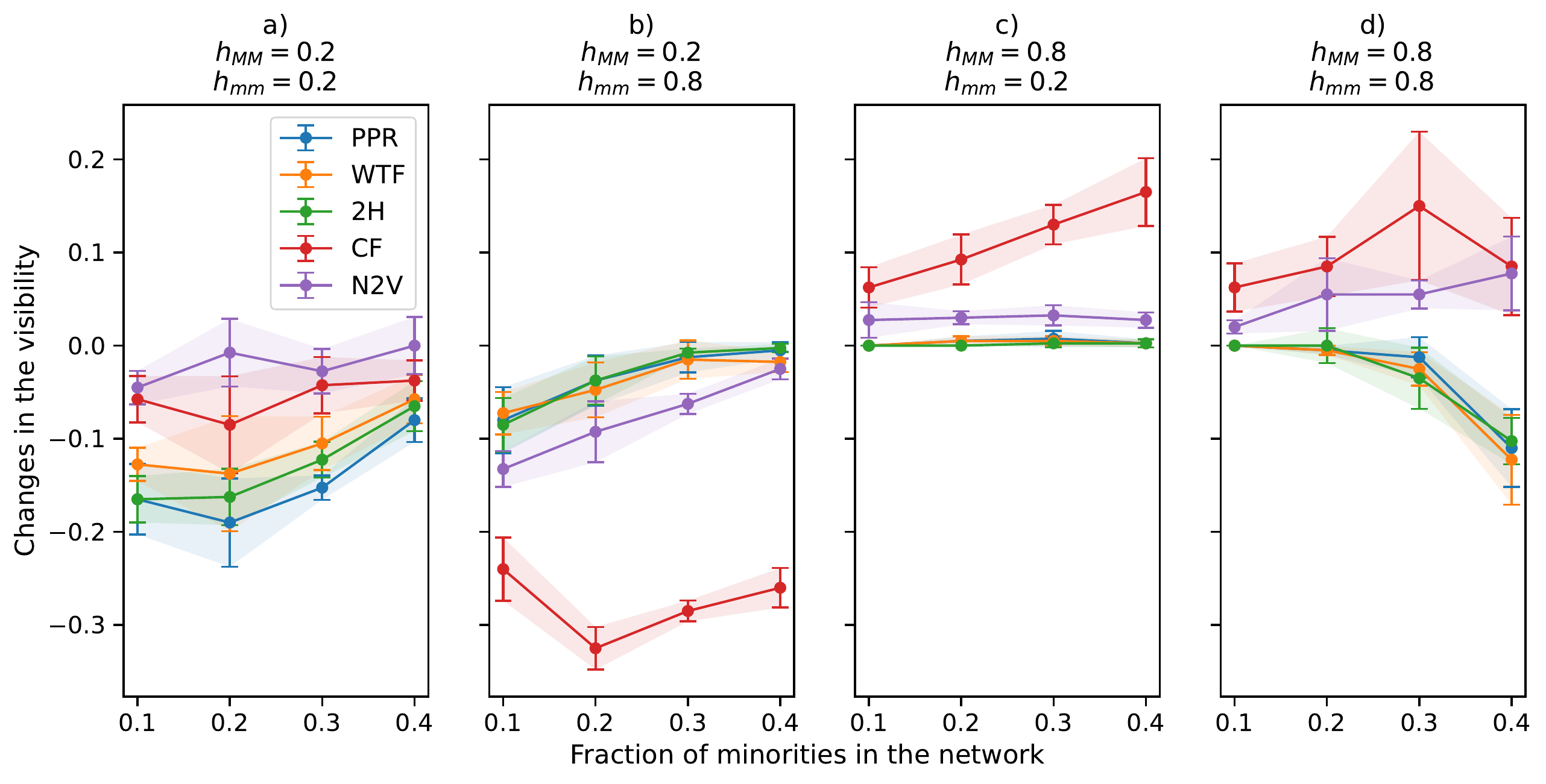}
    \caption{Changes in the visibility of minorities as a function of the minority size. The y-axis shows the change in visibility for the minority group after the recommendations. A positive (negative) change indicates that minorities appeared more (less) often in the top-10\% compared to their initial representation before the recommendations. If this change is around zero, the visibility of minorities remained constant or invariant. The x-axis shows the size of the minority group as a fraction of all nodes in the network. 
    In general, we see that larger minorities get penalized less than smaller ones when the majority is heterophilic (a,b). When the majority is homophilic, however, the changes in visibility not only depend on the fraction of the minority but also on its homophily level. For instance, when the minority is heterophilic (c), its visibility remains mostly constant for all algorithms except CF, and when the minority is homophilic (d), its visibility drops for larger-size minorities, unless N2V and CF are used. 
    %for different fraction of minorities in the network 
    }
    \label{fig:fm}
\end{figure*}

\subsection{RQ3: Is the change in visibility inversely proportional to the size of the minority or proportional to the in-group links within the minority?}
\para{Size of the minority:} To answer RQ1 and RQ2, we kept the size of the minority fixed ($f_m=0.3$) to study the effects of homophily on the visibility of minorities after the recommendations.
However, it is unclear whether the changes in visibility are inversely proportional to the size of the minority (e.g., larger changes for smaller minorities), or whether these are steady-state changes that appear regardless of the size of the minority.
In Figure~\ref{fig:fm}, we show how the change in visibility (y-axis) is affected by multiple factors including the size of the minority. 
First, we see a concordance among algorithms when the majority is heterophilic, Figures~\ref{fig:fm}(a) and~\ref{fig:fm}(b). In these cases, the larger the minority, the smaller the change in the visibility of the minority, except for CF which drastically reduces this visibility when the minority is more homophilic than the majority, Figure~\ref{fig:fm}(b). %the majority (b,e). %This can be explain by the fact that more links within the majority group are being created through the recommendations (see CF in~\Cref{fig:InterLinks} b).
When only the majority is homophilic, Figure~\ref{fig:fm}(c), i.e., most out-links point to nodes in the majority group, the size of minorities has almost no effect on their final visibility in algorithmic rankings unless CF is used as recommendation algorithm. When both groups are homophilic, Figure~\ref{fig:fm}(d), however, only CF and N2V increase the visibility of larger minorities more than the visibility of smaller minorities.

%In homophilic regimes, the size of minorities has almost no effect on their final visibility in algorithmic rankings unless CF or N2V are used as recommendations algorithms. These two algorithms increase the visibility of larger minorities more than the visibility of smaller minorities.

%Second, when the majority is homophilic (c,d), group size plays just a minor role in the visibility of the minority for PPR, WTF and 2H, while CF and N2V increases it. 
%Finally, when both groups are homophilic and the minority is more homophilic than the majority (e), we see that larger minorities loose visibility through CF and N2V, while they gain visibility (positive change) through the other algorithms.
%In summary, larger minorities get penalized less than smaller ones when the majority is heterophilic, and larger minorities benefit from the algorithms when they are homophilic.
%Note that throughout our experiments, CF and N2V have shown similar behaviors in terms of changes in network structure and visibility of minorities. However, they differ when we control for group size and the minority is homophilic and the majority heterophilic (see~\Cref{fig:fm} b).

\begin{figure*}[t!]
    \centering
    \includegraphics[width=0.99\textwidth]{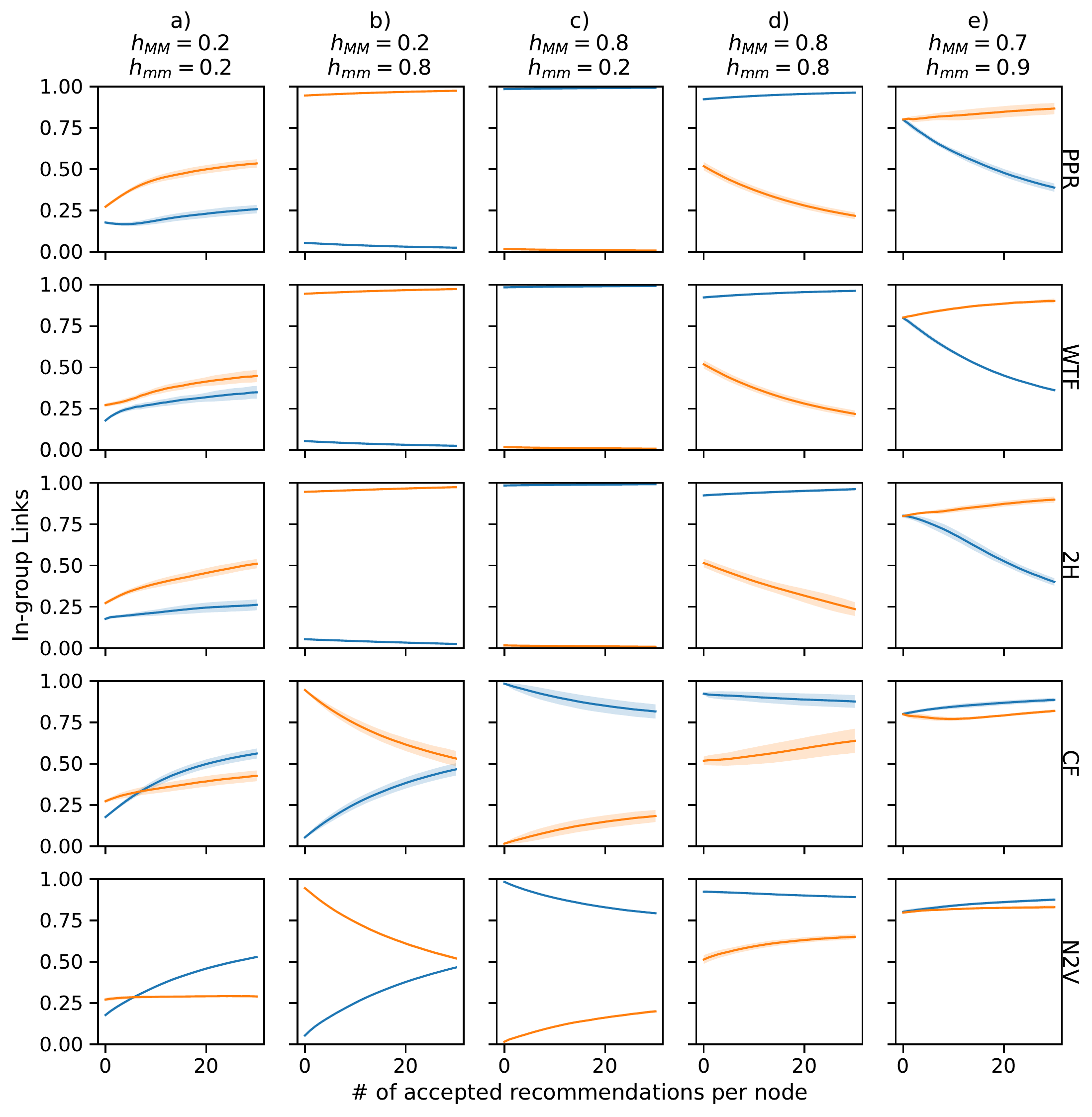}
    \caption{The evolution of the in-group links within majority nodes (blue) and within minority nodes (orange) for different recommendation algorithms (rows) and different values of homophily (columns) in the initial network. These networks posses a fixed number of nodes and fraction of minorities $f_m=0.3$.
    %At first glance we see two groups of results. The ones produced by PPR, WTF and 2H, and the ones by CF and N2V. 
    %Algorithms in each group change the in-group links in a similar fashion in each type of network. 
    % PPR, WTF and 2: These algorithms produce similar in-group links in each type of network. We see that when the majority is heterophilic (a,b) and when the minority is more homophilic than the majority (e), the in-group links within minorities increases through the recommendations. However, when minorities have higher homophily (b,e)
    }
    \label{fig:InterLinks}
\end{figure*}

% \para{What type of links are being added into the network?}
\para{In-group links:} 
As we have seen previously, the visibility of the minority can be affected by different factors, including the initial homophily of the network.
Since homophily depends on the mixing of types of edges (see~\cite{espin2022ineq} for a detailed derivation of homophily in DPAH networks), we further investigate the evolution of in-group links over time, see Figure~\ref{fig:InterLinks}. %(fraction of edges within the minority/majority group, see Section~\ref{sec:metrics}) 
%to explain our findings in Section~\ref{sec:rq2}. %Figure~\ref{fig:hv}.
Here, we found two main patterns. 
First, results from PPR, WTF and 2H are consistent in each type of network (columns). These algorithms mostly increase the number of in-group minority links, see Figures~\ref{fig:InterLinks}(a), \ref{fig:InterLinks}(b) and~\ref{fig:InterLinks}(e). 
Surprisingly, this advantage does not guarantee an increase in visibility for the minority group. On the contrary, they lose visibility, see Figures~\ref{fig:fm}(a) and~\ref{fig:fm}(b) for $f_m=0.3$.
Second, results from CF and N2V are also consistent in each type of network. We see in Figures~\ref{fig:InterLinks}(a) and ~\ref{fig:InterLinks}(b) that these two algorithms increase the in-group majority links when the majority is initially heterophilic, and reduce them when the majority is initially homophilic, see Figures~\ref{fig:InterLinks}(c), \ref{fig:InterLinks}(d) and \ref{fig:InterLinks}(e). 
% A similar effect occurs when both groups are homophilic (d). Although minorities loose in-group links, their vsibility is 
%. This means that by increasing the links within the minority, PPR, WTF and 2H, do not guarantee an
% \Cref{fig:hv_ppr,fig:hv_CF} 
%Figures 5(b) and 5(e)

% To explore how the connections between the different classes of nodes are changing, we analyzed the evolution of the fraction of links within the majority and within the minority, called in-group links in section [3.4].
Now, we analyze in details different possible homophily configurations.

When one class is homophilic and the other class is heterophilic, the links coming from both classes are mostly directed to nodes in the homophilic class. Let us consider PPR, WTF and 2H 
%b %-(PPR, WTF, 2H) and~\Cref{fig:InterLinks} c-(PPR, WTF, 2H), 
where the values of homophily are $h_{MM}=0.2$ for the majority and $h_{mm}=0.8$ for the minority and vice-versa, see Figures~\ref{fig:InterLinks}(b) and~\ref{fig:InterLinks}(c), respectively. In these situations, these recommendation algorithms will keep increasing the in-group proportions of the homophilic group since the recommended links mostly point to nodes in this group. % the connection between the classes, and the links will stay directed towards the class with the higher homophily value. 
These correspond to situations in the white regions at the top-left and bottom-right of Figures~\ref{fig:hv_ppr} to~\ref{fig:hv_2H}. Hence, this shows that the absence of variation in the fraction of minority is due to the fact that PPR, WTF and 2H do not modify connections between classes in these cases. This does not hold for CF and N2V. In fact, under the same homophily conditions, these methods make the in-group links for both classes more similar, decreasing structural differences between classes, see Figures~\ref{fig:InterLinks}(b) and ~\ref{fig:InterLinks}(c). % b,c-(CF, N2V). 

Now, we will consider regimes where both classes are heterophilic, $h_{mm}=h_{MM}=0.2$, see Figure~\ref{fig:InterLinks}(a). Here, CF and N2V are the only algorithms in which the initial conditions of in-group links are flipped. Note that the proportion of links within the majority group gets larger than the proportion of links within the minority after multiple rounds of recommendations. Consequently, the majority increases its visibility even further by pushing minorities to lower ranks, see Figures~\ref{fig:hv_CF} and~\ref{fig:hv_N2V}.
% and the fraction of in-group links for the majority becomes more than the fraction of in-group links for the majority. 
Interestingly, the visibility of minorities decreases even if the flip does not occur in these heterophilic settings for PPR, WTF and 2H, see %, the visibility of minority got reduced (
bottom-left of Figures~\ref{fig:hv_ppr} to ~\ref{fig:hv_2H}.
%, ~\Cref{fig:hv_wtf} and ~\Cref{fig:hv_2H}).

On the other extreme of homophily, when both groups are homophilic, $h_{mm}=h_{MM}=0.8$, we found two main patterns, see Figure~\ref{fig:InterLinks}(d).
% Instead, we analyze now the symmetric homophily regime, which is when the two classes are homophilic at the same level as in Figure~\ref{fig:InterLinks} d), where the level of homophily is $0.8$ for both classes. 
%In the initial configuration, the two classes have the same value of homophily and nodes tend to connect within the same class. But the presence of preferential attachment combined with high homophily levels leads to the formation of hubs (nodes with high indegree) among the majority class. 
First, PPR, WTF and 2H tend to strengthen the connections towards the majority group by either recommending majority-to-majority or minority-to-majority links. This in turn penalizes the minorities at the top of the rank, see $h_{mm}=h_{MM}=0.8$ in Figures ~\ref{fig:hv_ppr} to ~\ref{fig:hv_2H}.
% First, PPR, WTF and 2H tend to strengthen the connection within majorities and will force part of the nodes belonging to the minority to link to majority nodes. This results in penalization of minorities at the top of the rankings (See the blue squares in top-right of Figures ~\ref{fig:hv_ppr} to ~\ref{fig:hv_2H}).
In contrast, CF and N2V slowly increase the number of connections within the minority group. For N2V, one possible explanation is that the homophily levels are high enough so that the two classes (especially the minority class), are represented in the embeddings as, at least partially, separated clusters.
% the effect is instead different for N2V and CF. Here, their tendency is to slowly increase connections between minorities. For N2V, one possible intuitive explanation is that the homophily levels are high enough so that the two classes (and specifically the minority class), are represented in the embeddings as, at least partially, separated clusters.

Lastly, the possibility to systematically vary the initial levels of homophily for both classes allows us to identify tipping points. 
For instance, in a homophilic regime, where both groups have the same level of initial homophily, $h_{MM}=h_{mm}=0.8$, we found that PPR, WTF and 2H increase the number of links within the majority group after multiple recommendations, see Figure~\ref{fig:InterLinks}(d). However, the same algorithms may also increase the number of links within the minorities, and thus their visibility, if the minority group is initially more homophilic than the majority, $h_{MM}=0.7$ and $h_{mm}=0.9$, see Figure~\ref{fig:InterLinks}(e).
% An example is shown for PPR in Figure~\ref{fig:InterLinks}(e), for values of homophily $h_{MM}=0.7$ and $h_{mm}=0.9$ for the majority and the minority classes, respectively. 
% We have seen how in the case of symmetric homophilic networks, see PPR, WTF, 2H in Figure~\ref{fig:InterLinks}(d), majority nodes were increasing the connections among themselves. 
% Instead, PPR in Figure~\ref{fig:InterLinks}(e), with a small modification of the initial homophily values, minorities are able to better increment their connections, resulting in an increase in their visibility. 
CF and N2V, on the other hand, do not show this tipping effect when both groups of nodes are initially homophilic. In either case, these two algorithms keep increasing the proportion of in-group links which induces segregation. %, see see Figures~\ref{fig:InterLinks}(d) and~\ref{fig:InterLinks}(e).
% CF and 2H, instead, separate majorities and minorities even more (both classes have high in-group links), as they did when the homophily was $0.8$ for both classes.
% for each algorithm (colors) and type of network (columns) how t

% \begin{comment}

% \lesp{Instead of trying different top-k's, maybe it is better to show for each recommender system at which top-k they reach for the first time statistical parity. Then we can say recsys1 is the fairest among all the ones we try.}

% xxx.

% \para{Top-k in-degree:}
% xxx.

% \para{Top-k PageRank:}
% x...

% \begin{figure}[h]
%     \centering
%     \subfigure[PPR]{\includegraphics[width=0.23\textwidth]{PPR-DPAH-N1000-fm0.3-d0.03-ploM2.5-plom2.5_heatmap_diff_minority.png} \label{fig:synhom-pr}}
%     \subfigure[Two-Hops]{\includegraphics[width=0.23\textwidth]{Twohops-DPAH-N1000-fm0.3-d0.03-ploM2.5-plom2.5_heatmap_diff_minority.png} \label{fig:synhom-2h}}
%     \subfigure[Node2Vec]{\includegraphics[width=0.23\textwidth]{Node2vec-DPAH-N1000-fm0.3-d0.03-ploM2.5-plom2.5_heatmap_diff_minority.png} \label{fig:synhom-n2v}}
%     \subfigure[CF]{\includegraphics[width=0.23\textwidth]{CF-DPAH-N1000-fm0.3-d0.03-ploM2.5-plom2.5_heatmap_diff_minority.png} \label{fig:synhom-cf}}
%     \caption{\textbf{Interlinks.}  Explain}
%     \label{fig:synhom}
% \end{figure}

% \end{comment}

%%%%%%%%%%%%%%%%%%%%%%%%%%%%%%%%%%%%%%%%%%%%%%%%%%%%%%%%%
% EMPIRICAL
%%%%%%%%%%%%%%%%%%%%%%%%%%%%%%%%%%%%%%%%%%%%%%%%%%%%%%%%%

%\section{Empirical Evidence}

%We analyze x real-world directed networks. \Cref{tb:empirical} shows xx

\section{Limitations and future work}
We have limited our study to five recommendation algorithms, and in future work we aim to include more algorithms into this investigation, especially recent versions of popular algorithms that have been developed with the goal to increase fairness. 

Furthermore, we focused on scale-free directed networks with homophily which represent a plausible configuration of online social networks. As next steps, we would like to include in our analysis different network simulation models that include other factors in the network generation process, such as multiple node-attributes, heterogeneous group mixing, the presence of communities, and triadic closure. We also acknowledge the fact that our analysis is theoretical and has not been validated with real data. We plan to extend our study by considering empirical networks.

Importantly, link recommendation algorithms and datasets are generally proprietary. This is why simulation-based approaches are often necessary for this kind of investigations. In addition, the simulation approach enables us to examine different scenarios which might not occur in one instance of the data~\cite{steinbacher2021advances}. 

% Lastly, further research is needed to study the impact of recommendation algorithms on networks with heterogeneous group mixing and for networks with more than two groups.

%For future work it would be important to empirically investigate how different link recommendations are performed on social media platforms and which ones are more effective. 

\section{Conclusions}
%lets summerize main findings/take aways here
In this work, we systematically studied five link recommendation algorithms and quantified their feedback loop effects on bi-populated scale-free directed networks with homophily. In particular, we assessed two types of changes in these networks due to multiple link recommendations. First, we measured the changes in network structure in terms of clustering and in-degree distribution. Second, we measured the changes in the visibility of minorities at the top-10\% of the rank with respect to their PageRank (importance) scores, highlighting the effects of homophily, minority size, and in-group links.

Our results show that four out of the five algorithms reduced on average the visibility of minorities more often than to the majority counterpart. 
In particular, PPR, WTF and 2H when both groups are initially heterophilic, and CF when the minority is initially more homophilic than the majority. %the initial homophily of the minority is larger that the initial homophily of the majority.
%Interestingly, both CF and N2V show more symmetric changes in visibility for both the minority and the majority group across all combinations of group homophily.
% Our results show that 4 out of 5 algorithms reduced on average the visibility of minorities more than the visibility of majorities in scale-free social networks with various levels of homophily. 
% Especially, for PPR, WTF and 2H one can see that the minority loses more visibility than the majority (especially in the heterophilic regime), while CF and N2V show more symmetric effects on the visibility of the minority and majority.

We also found that while all algorithms tend to close triangles and increase the clustering coefficient, all algorithms except N2V are prone to favor and suggest nodes with high in-degree. This is known as popularity bias, rich-get-richer effect or cumulative advantage, a well-known mechanism that contributes to inequality. 
% To better understand why CF and N2V are different, we explored their link recommendation behavior. While all algorithms tend to close triangles and increase the clustering coefficient, all algorithms except N2V are prone to favor and suggest nodes with high in-degree. This is known as the rich-get-richer effect or cumulative advantage, a well-known mechanism that contributes to inequality. 

Link recommendations based on N2V rely on the proximity of nodes in the embedding space, which does not necessarily imply closeness to nodes with high in-degree. Consequently, N2V is a promising alternative to other link recommendation algorithms since it mitigates cumulative advantage.
% Link recommendations of N2V are based on closeness in the embedding space and similar nodes are close to each other in the embedding but they are not necessarily close to nodes with high in-degree. Consequently, N2V is a promising alternative to other link recommendation algorithm that helps to avoid exaggerating cumulative advantage effects with each iteration.

\begin{acks}
This work has received funding by the European Union's Horizon 2020 research and innovation programme under the Marie Sk\l{}odowska-Curie Actions (grant agreement number 860630) for the project : ``NoBIAS - Artificial Intelligence without Bias''. Furthermore, this work reflects only the authors' view and the European Research Executive Agency (REA) is not responsible for any use that may be made of the information it contains. Moreover, Fariba Karimi was supported by the Austrian research agency (FFG) (project number 873927).
\end{acks}

\balance
\bibliographystyle{ACM-Reference-Format}
\bibliography{main}

%\appendix

\end{document}